\documentclass[journal]{IEEEtran}

\usepackage{graphicx}
\usepackage{amsmath}
\usepackage{mathtools}
\usepackage[makeroom]{cancel}
\usepackage{amssymb}
\usepackage{comment}
\usepackage{float}
\usepackage[dvipsnames]{xcolor}
\usepackage{ifpdf}
\usepackage{cite}
\usepackage{hyperref}
\usepackage{enumitem}
\usepackage{algorithm2e}
\usepackage{bbold}
\usepackage{gensymb}
\usepackage{multicol}
\begin{document}

\title{Using Effective Generator Impedance for Forced Oscillation Source Location}

\author{Samuel Chevalier,~\IEEEmembership{Student Member,~IEEE,}
        Petr Vorobev,~\IEEEmembership{Member,~IEEE,}
        Konstantin Turitsyn,~\IEEEmembership{Member,~IEEE}
\thanks{This work was supported in part by the MIT/Skoltech initiative, the Skoltech-MIT Next Generation grant, and the MIT Energy Initiative Seed Fund Program.}

\thanks{S. Chevalier and K. Turitsyn are with Department of Mechanical Engineering, Massachusetts Institute of Technology. E-mail: schev, turitsyn@mit.edu

P. Vorobev is with Department of Mechanical Engineering, Massachusetts Institute of Technology and also with Skolkovo Institute of Science and Technology. E-mail: petrvoro@mit.edu }}

\maketitle

\begin{abstract}
Locating the sources of forced low-frequency oscillations in power systems is an important problem. A number of proposed methods demonstrate their practical usefulness, but many of them rely on strong modeling assumptions and provide poor performance in certain cases for reasons still not well understood. This paper proposes a systematic method for locating the source of a forced oscillation by considering a generator's response to fluctuations of its terminal voltages and currents. It is shown that a generator can be represented as an effective admittance matrix with respect to low-frequency oscillations, and an explicit form for this matrix, for various generator models, is derived. Furthermore, it is shown that a source generator, in addition to its effective  admittance, is characterized by the presence of an effective current source thus giving a natural qualitative distinction between source and non-source generators. Detailed descriptions are given of a source detection procedure based on this developed representation, and the method's effectiveness is confirmed by simulations on the recommended testbeds (eg. WECC 179-bus system). This method is free of strong modeling assumptions and is also shown to be robust in the presence of measurement noise and generator parameter uncertainty.
\end{abstract}


\IEEEpeerreviewmaketitle


\section{Introduction}\label{Introduction}

\IEEEPARstart{W}{ith} the recent widescale deployment of Phasor Measurement Units (PMUs) across the US transmission grid~\cite{Achanta:2017}, system operators are becoming keenly aware of the pervasive presence of low frequency oscillations. Generally, low frequency oscillations are either natural modes, attributed to poorly tuned control settings and large power flows across weak tie lines, or forced oscillations, which are caused by extraneous disturbances. Such extraneous inputs may be related to faulty controllers, turbine vibrations, or cyclical loads~\cite{Vanfretti:2012,Rostamkolai:1994,Wang:2010}. The appearance of forced oscillations reduces the quality of electric power and has potential detrimental effects on various equipment. More importantly, whenever a disturbance occurs at the frequencies close to one of the natural system modes, a resonance condition may lead to significant amplification of amplitude, where a relatively small perturbation on one bus can cause rather large power swings in different locations around the system. An example of this effect is the 2005 WECC disturbance where a reasonably small 20MW oscillation at the Nova Joffre cogeneration power plant in Canada resonated with one of the main inter-area modes resulting in a 200MW power oscillation on the Oregon-California intertie~\cite{Nezam:2016}. 

Accordingly, there is a need in the power systems community for the development of methods which are capable of using on-line PMU data to trace the source of a forced oscillation. It is accepted that designing control methods for damping of forced oscillations is impractical~\cite{Maslennikov:2017}; instead, disconnection of the identified source with subsequent investigation of the causes of the disturbance is the main solution. A variety of source identification techniques have been developed with varying levels of success; many are outlined in a recent literature survey~\cite{Wang:2017} where the main requirements for such methods are also stated. A set of test cases for validating different source location methods is presented in \cite{Maslennikov:2016}. These cases were developed in coordination with IEEE Task Force on Forced Oscillations, and they will allow for a standardized examination of all source detection algorithms.

Before applying any source location procedure, the type of observed disturbance has to be identified. To differentiate between forced oscillations and other types of disturbances, a method based on statistical signatures of different types of oscillations was proposed in ~\cite{Wang:2016}. Similarly, ~\cite{OBrien:2017} uses spectral analysis of PMU data to ``trigger" a forced oscillation warning. The authors then suggest using statistical tools (pattern mining and maximal variance ratios) from on-line generator SCADA data to determine the oscillation source. If oscillation magnitudes are low and signal noise is high, \cite{Zhou:2015} proposes using the self-coherence spectrum of a PMU signal and its time shifted version to perform forced oscillation detection. In~\cite{Wilson:2014}, phase coherency is used to identify groups of generators which swing together. The source is identified as the generator in the source group which is providing the smallest contribution to the overall damping. This will correspond to the generator whose rotor oscillation phase is leading all other source group rotor oscillation phases.

An important class of source location methods, which are termed the hybrid methods in~\cite{Wang:2017}, leverage both a known system model and measured PMU data. Demonstrated in~\cite{Wu:2012} and~\cite{Ma:2010}, these methods use measured PMU signals as inputs for a power system model. After simulating this model, the time domain model outputs are compared with their corresponding measured PMU signals. Significant deviation between the model predictions and the PMU measurements may indicate the presence of a forced oscillation. These types of methods are also used for model validation.

One of the most promising methods, which has already shown its practical performance, is the  Transient Energy Flow (TEF) method, initially developed in~\cite{Chen:2013}. One of the main advantages of this method is that it tracks the flow of effective transient energy in all lines where PMU data is available, thus being naturally model independent. The authors show that the dissipated energy is equivalent to damping torque. Adaption of the method for use with actual PMU data, as outlined in~\cite{Maslennikov:2017}, has been named the Dissipating Energy Flow (DEF). The method  was able to successfully locate the source of a forced oscillation in a variety of simulated test cases and in over 50 actual events from both ISONE and WECC. 

While having the advantage of being model free, this method has certain shortcomings, the most important being its inability to distinguish between a true source bus and a bus having an effective ``negative damping" contribution, since both such buses are seen as sources of Transient Energy. A number of rather strong assumptions are also crucial for the method, namely, constant PQ loads and a lossless network. Accordingly, the method performs poorly when constant impedance loads are present. In this situation, the method triggers a ``false alarm"~\cite{Wang:2017} by identifying such a load as the disturbance source. This particular shortcoming raises a natural question about the proper definition of oscillation energy. A full discussion of the open questions concerning the DEF/TEF methods can be found in the conclusion of~\cite{Chen:2017}.

It is clear that a more systematic approach is needed to study forced oscillations, especially in the development of methods which do not heavily rely on strong model assumptions. In this paper we have developed a systematic procedure to locate the sources of forced oscillations. We start by deriving a relation between generator terminal voltage and current fluctuations in the presence of persistent oscillation. We then show with minimal modeling assumptions that, based on this relation, it is possible to effectively distinguish between source and non-source generators. We also apply our results to perturbations with frequencies close to a natural system mode so that the maximum amplitude is observed on a non-source generator. The specific contributions of this paper are as follows.

\begin{enumerate}
\item A systematic method for calculating a generator's frequency response function, with respect to terminal voltage and current perturbations, is given.

\item An \emph{equivalent circuit} interpretation is introduced which treats a generator's frequency response function as an effective admittance matrix $\mathcal{Y}$ and any internal forced oscillations as current injections $\mathcal{I}$.

\item  An explicit forced oscillation source location algorithm, which compares predicted and measured current spectrums while making unique measurement noise considerations, is presented.
\end{enumerate}

The rest of the paper is structured as follows. In Section \ref{Generator Modeling} we introduce an effective generator admittance matrix $\mathcal{Y}$ with respect to terminal voltage and current perturbations and show that a forced oscillation source may be transformed into an effective current injection $\mathcal{I}$. We show the explicit steps for building the admittance matrix and current injections associated with a classical generator, and we then extend the methods to a $6^{\rm th}$ order generator with voltage control. In Section \ref{Method}, we present an algorithm for using $\mathcal{Y}$ to determine if a generator is the source of an oscillation. Section \ref{Test_Results} presents test results from a 3-bus system and from the standardized 179-bus test cases of~\cite{Maslennikov:2016} in the presence of measurement noise and generator parameter uncertainty. Also, we include a comparison between our algorithm and the DEF method in the context of a system with a resistive load. Finally, conclusions and plans for future work are offered in Section \ref{Conclusion}.


\section{Representing Generators as Frequency Response Functions}\label{Generator Modeling}
This section introduces the concept of a generator's effective admittance matrix $\mathcal{Y}$ which characterizes its frequency response. If the generator is an oscillatory source, then in addition to matrix $\mathcal{Y}$, we show that an effective current source $\mathcal{I}$ will appear in parallel with admittance $\mathcal{Y}$. We analytically derive these expressions for a classical generator model and then show how the methods extend to higher order models.

\subsection{State Space Formulation for a Classical Generator}\label{Class_Gen}
\begin{figure}
\begin{centering}
\includegraphics{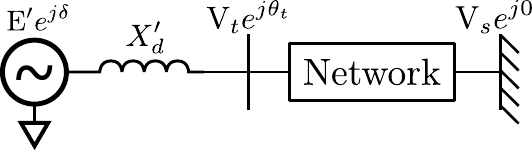} 
\par\end{centering}
\caption{\label{fig: 2_Order_Gen} $2^{{\rm nd}}$ order generator tied to a network. Internal generator voltage ${\rm E}'e^{j\delta}$, terminal voltage ${\rm V}_te^{j\theta_t}$, and swing bus voltage ${\rm V}_se^{j\theta_s}$ with $\theta_s=0$ are all shown.}
\end{figure}

\begin{figure}
\begin{centering}
\includegraphics[scale=0.9]{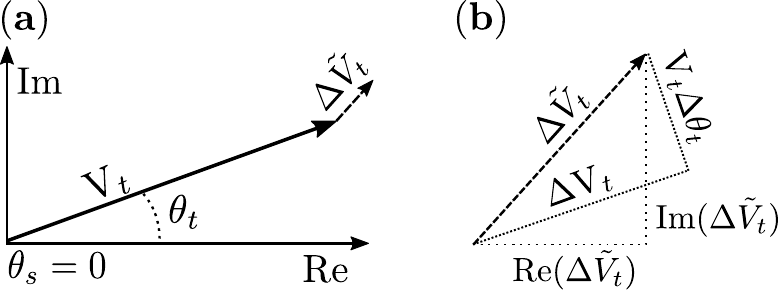} 
\par\end{centering}
\caption{\label{fig: Polar_Rect_Relationships} Panel $({\bf a})$ shows the steady state phasor ${\rm V}_te^{j\theta_t}$ and phasor deviation $\Delta{\tilde V}_t$. Panel $({\bf b})$ expands deviation $\Delta{\tilde V}_t$ from panel $({\bf a})$ and decomposes the relationship between the rectangular deviations (${\rm Re}(\Delta {\tilde V}_t)$, ${\rm Im}(\Delta {\tilde V}_t)$) and the corresponding polar deviations ($\Delta {\rm V}_t$, ${\rm V}_t \Delta {\theta_t}$).}
\end{figure}

In this section, the admittance matrix which relates a classical generator's rectangular voltage and rectangular current perturbations is derived. Effective current sources relating to torque and EMF oscillations are also derived. We start by considering a $2^{\rm nd}$ order generator with its internal EMF magnitude fixed. This generator is connected to some terminal bus with positive sequence phasor voltage ${\rm V}_te^{j\theta_t}$ at frequency $\omega_0$. This configuration is shown by Fig. \ref{fig: 2_Order_Gen}.
In order to quantify the admittance matrix ($\mathcal{Y}$) and current injection ($\mathcal{I}$) associated with this generator, a linearized state space formulation is used.
\begin{align}
\Delta{\dot {\bf x}} &= A\Delta{\bf x} + B\Delta{\bf u}\\
\Delta{\bf y}  &= C\Delta{\bf x} + D\Delta{\bf u},
\end{align}
where the state variable vector ${\bf x}$ contains the torque angle ($\delta$) and speed deviation ($\Delta\omega$) of the generator, and the input vector ${\bf u}$ contains the mechanical torque variations, two orthogonal terminal bus voltages, and the generator EMF. These are expressed as
\begin{align}
{\bf x} &= \left[\delta\;\;\Delta\omega\right]^{\top}\\
{\bf u}\label{eq: u} &= \left[\tau_{m}\;\;{\rm Re}(\tilde{V}_{t})\;\;{\rm Im}(\tilde{V}_{t})\;\;{\rm E}'\right]^{\top}.
\end{align}
The swing equation for the $2^{\rm nd}$ order generator is formulated with polar variables using a quasi-stationary power flow approximation. We neglect armature resistance $R_a$ since it is typically $\sim 1\%$ of transient reactance $X'_d$.
\begin{align}
\dot{\delta} &= \Delta \omega\\
\label{eq: swing}M\Delta\dot{\omega} &= \tau_{\rm m} - \frac{{\rm V}_t{\rm E}'}{X'_d}\sin(\delta - \theta_t) - D\Delta\omega,
\end{align}
where in this expression, we have also assumed $P_m=\omega\tau_m\approx\tau_m$ since the speed deviations are small. This expression may be linearized and expressed in state space formulation. $\Delta{\bf u}_{V_p}$ is the input vector of polar voltage perturbations, $\Delta u_\tau$ is the input torque perturbation, $\Delta u_{\rm E}$ is the input EMF variation, and power angle is defined as $\varphi=\delta-\theta_t$:
\begin{align}
\Delta{\bf \dot{x}}= & A\Delta{\bf x}+B_{V_p}\Delta{\bf u}_{V_p}+B_{\tau}\Delta u_{\tau}+B_{{\rm E}}\Delta u_{\rm E}\\
\left[\begin{array}{c}
\Delta\dot{\delta}\\
\Delta\dot{\omega}
\end{array}\right]= & \left[\begin{array}{cc}
0 & 1\\
-\frac{{\rm V}_{t}{\rm E}'}{MX'_{d}}\cos(\varphi) & -\frac{D}{M}
\end{array}\right]\left[\begin{array}{c}
\Delta\delta\\
\Delta\omega
\end{array}\right]+\\
&\left[\begin{array}{cc}
0 & 0\\
-\frac{{\rm E}'}{MX'_{d}}\sin(\varphi) & \frac{{\rm V}_{t}{\rm E}'}{MX'_{d}}\cos(\varphi)
\end{array}\right]\left[\begin{array}{c}
\Delta{\rm V}_{t}\\
\Delta\theta_{t}
\end{array}\right]+\nonumber\\
 & \left[\begin{array}{c}
0\\
\frac{1}{M}
\end{array}\right]\left[\begin{array}{c}
\Delta\tau_{m}\end{array}\right]+\left[\begin{array}{c}
0\\
\frac{-{\rm V}_{t}}{MX'_{d}}\sin\left(\varphi\right)
\end{array}\right]\left[\Delta{\rm E}^{'}\right]\nonumber.
\end{align}
In deriving this model, we wish to relate terminal voltage and current perturbations in rectangular coordinates. To do so, small perturbations of the voltage magnitude $\Delta {\rm V}_t$ and phase $\Delta \theta_t$ on the terminal bus voltage ${\tilde V}_t$ are considered, such that
\begin{align}
{\tilde V}_t + \Delta {\tilde V}_t =& ({\rm V}_t + \Delta {\rm V}_t)e^{j(\theta_t+\Delta \theta_t)}\label{eq: delta_Vt}.
\end{align}
After linearizing, the $\Delta {\tilde V}_t$ components may be separated into their real and imaginary parts, and the polar rectangular relationships may be expressed by employing transformation matrix ${\rm T}_1$. Fig. \ref{fig: Polar_Rect_Relationships} graphically portrays the following relationships:
\begin{align}\label{eq: T_1}
\left[\begin{array}{c}
{\rm Re}(\Delta {\tilde V}_{t})\\
{\rm Im}(\Delta {\tilde V}_{t})
\end{array}\right] & =\left[\begin{array}{cc}
\cos(\theta_{t}) & -{\rm V}_t\sin(\theta_{t})\\
\sin(\theta_{t}) & {\rm V}_t\cos(\theta_{t})
\end{array}\right]\left[\begin{array}{c}
\Delta{\rm V}_{t}\\
\Delta\theta_{t}
\end{array}\right]\\
\Delta{\bf u}_{V_r} & =\left[{\rm T}_1\right]\Delta{\bf u}_{V_p}.\nonumber
\end{align}
Accordingly, the inverse transformation matrix ${\rm T}_1^{-1}$ from (\ref{eq: T_1}) is employed to transform the vector of polar voltage perturbation variables ($\Delta{\bf u}_{V_p}$) into the vector of rectangular voltage perturbation variables ($\Delta{\bf u}_{V_r}$). The corresponding state space matrix is $B_{V_r}$, where $B_{V_r}=B_{V_{p}}{\rm T}_1^{-1}$. This is used to reformulate the system's state space representation: 
\begin{align}
\Delta{\bf \dot{x}}\label{eq: p2r}= & A\Delta{\bf x}+B_{V_{r}}\Delta{\bf u}_{V_{r}}+B_{\tau}\Delta u_{\tau}+B_{{\rm E}}\Delta u_{\rm E}.
\end{align}
$B_{V_r}$ has the following analytical structure:
\begin{align}
B_{V_{r}} & =\frac{{\rm E}'}{MX'_{d}}\left[\begin{array}{cc}
0 & 0\\
-\sin(\delta) & \cos\left(\delta\right)
\end{array}\right].
\end{align}
The state space model's output ${\bf y}$ is defined as the orthogonal real and imaginary current flows {\bf into} the generator (we call these the \textit{negative current injections}):
\begin{align}
I & =\frac{\left[{\rm Re}({\tilde V}_{t})+j{\rm Im}({\tilde V}_{t})\right]-{\rm E}'e^{j\delta}}{jX'_{d}}.
\end{align}
$I$ is linearized and split into real and imaginary currents.
\begin{align}\Delta{\bf y}= & C\Delta{\bf x}+D_{V_{r}}\Delta{\bf u}_{V_{r}}+D_{{\rm E}}\Delta u_{{\rm E}}\\
\left[\begin{array}{c}
\Delta{\rm Re}\left(I\right)\\
\Delta{\rm Im}\left(I\right)
\end{array}\right]= & \left[\begin{array}{cc}
-\frac{{\rm E}'\cos\left(\delta_{0}\right)}{X'_{d}} & 0\\
-\frac{{\rm E}'\sin\left(\delta_{0}\right)}{{X'_{d}}} & 0
\end{array}\right]\left[\begin{array}{c}
\Delta\delta\\
\Delta\omega
\end{array}\right]+\\
\left[\begin{array}{cc}
0 & \frac{1}{X'_{d}}\\
-\frac{1}{X'_{d}} & 0
\end{array}\right] & \left[\begin{array}{c}
\Delta{\rm Re}\left(V_{t}\right)\\
\Delta{\rm Im}\left(V_{t}\right)
\end{array}\right]+\frac{1}{X'_d}\left[\begin{array}{c}
-\sin(\delta)\\
\cos(\delta)
\end{array}\right]\left[\Delta{\rm E}'\right]\nonumber
\end{align}
\subsection{Frequency Response Function Construction}
With the state space model formulated, the Fourier transform of the system may be taken, such that $\dot{\bf x} = j\Omega_d {\bf \tilde x}$. In this analysis, we note that ${\tilde u}_{\rm E}={\tilde {\rm E}'}$ and ${\tilde u}_{\tau}={\tilde \tau}$ are the Fourier transforms of oscillatory steady state deviations, where the respective steady state values are given by ${\rm E}'$ and $\tau_0$.
\begin{align}
j\Omega_d {\bf \tilde x} &= A\tilde{\bf x}+B_{V_r}\tilde{\bf u}_{V_r}+B_{\tau}\tilde{ u}_{\tau}+B_{\rm E}{\tilde u}_{\rm E}\\
\tilde{\bf y}          &= C\tilde{\bf x}+D_{V_r}\tilde{\bf u}_{V_r}+D_{{\rm E}}\tilde{u}_{{\rm E}}.
\end{align}
The Frequency Response Functions (FRFs), which directly relate the inputs to the outputs, can be solved for, where $\Theta=(j\Omega_d {\mathbb 1} - A)^{-1}$:
\begin{align}
\tilde{\bf x} =& \Theta(B_{V_r}\tilde{\bf u}_{V_r}+B_{\tau}\tilde{ u}_{\tau}+B_{\rm E}{\tilde u}_{\rm E})\\
& \qquad\qquad\;\;\Downarrow \nonumber\\
\tilde{\bf y}    =& \left [C\Theta B_{V_r}+D_{V_r} \right ] \tilde{\bf u}_{V_r}+ \\
                  & \left [ C\Theta B_{\tau} \right ] \tilde{ u}_{\tau}+\left [ C\Theta B_{\rm E} +D_{{\rm E}} \right ] \tilde{u}_{\rm E}\nonumber.
\end{align}
In this formulation, the following observations may be made. The  FRF which relates terminal bus voltage differentials to the current flows acts as an admittance matrix. Similarly, the FRF relating the torque phasor to the currents flows, in conjunction with the torque phasor, acts as one \textit{potential} current source, and the FRF relating the generator EMF phasor to the currents flows, in conjunction with the EMF phasor, acts as a second \textit{potential} current source:
\begin{align}
{\mathcal Y} &= C(j\Omega_d {\mathbb 1} - A)^{-1}B_{V_r}+D_{V_r} \label{eq: Ymc}\\
{\mathcal I}_\tau &= \left [ C(j\Omega_d {\mathbb 1} - A)^{-1}B_{\tau} \right ] \left[\tilde \tau_m \right]\label{eq: Imc_tau}\\
{\mathcal I}_{\rm E} &= \left [ C(j\Omega_d {\mathbb 1} - A)^{-1}B_{\rm E} +D_{{\rm E}}\right ] [{\tilde {\rm E}}' ]\label{eq: Imc_E}.
\end{align}
With this observation, the following intuitive model formulation may be observed:
\begin{equation}
\label{eq: I_Flow}\left[\begin{array}{c}
\tilde{I}_R\\
\tilde{I}_I
\end{array}\right] = {\mathcal Y} \left[\begin{array}{c}
\tilde{V}_{R}\\ 
\tilde{V}_{I}
\end{array}\right] + {\mathcal I}_\tau + {\mathcal I}_{\rm E},
\end{equation}
where ${\mathcal Y}$ is a $2\times2$ matrix and the real (or imaginary) part of the voltage (or current), which is itself a phasor, is given by $\tilde{V}_{R}$. The structure of $\mathcal{Y}$ may be written explicitly as
\begin{align}
\label{eq: Y}\mathcal{Y} & =\Gamma\left[\begin{array}{cc}
\sin\delta\cos\delta & -\cos^{2}\delta\\
\sin^{2}\delta & -\sin\delta\cos\delta
\end{array}\right]+\left[\begin{array}{cc}
0 & \frac{1}{X'_{d}}\\
\frac{-1}{X'_{d}} & 0
\end{array}\right]\\
\Gamma & =\frac{\frac{{\rm E}'^{2}}{X'^{2}_{d}}}{\left(\frac{{\rm V}_{t}{\rm E}'}{X'_{d}}\cos(\varphi)-M\Omega_{d}^{2}\right)+j\left(\Omega_{d}D\right)}
\end{align}
and the negative current injections $\mathcal{I}_\tau$ and $\mathcal{I}_{\rm E}$ are given as
\begin{align}
\mathcal{I}_{\tau} & =-\Gamma\frac{X'_{d}}{{\rm E}'}\left[\begin{array}{c}
\cos(\delta)\label{eq: I_t}\\
\sin(\delta)
\end{array}\right]\tilde{\tau}_{m}\\
\mathcal{I}_{{\rm E}} & =\left (\Gamma\frac{{\rm V}_t\sin\left(\varphi\right)}{{\rm E}'}\left[\begin{array}{c}
\cos(\delta)\\
\sin(\delta)
\end{array}\right]+\left[\begin{array}{c}
-\frac{\sin(\delta)}{X'_d}\\
\frac{\cos(\delta)}{X'_d}
\end{array}\right]\right)\tilde{{\rm E}}'\label{eq: I_e}.
\end{align}
When a generator is the source of negative damping, the angle associated with the complex admittance matrix parameter $\Gamma$ will point into quadrants I or II of the complex plane. Accordingly, the FRF of a generator provides a natural interpretation of negative damping with regards to the phase shift relationships between the input and output signals. Future work shall investigate how this property may be exploited to find locations of negative damping in the system.
\subsection{Transformation to a Local dq Reference Frame}
\begin{figure}
\begin{centering}
\includegraphics[scale=0.85]{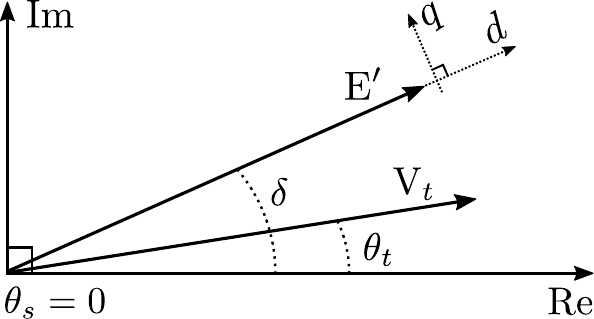}
\par\end{centering}
\caption{\label{fig: DQ_Qxis} Orientation of the direct ($d$) and quadrature ($q$) axes.}
\end{figure}

\begin{figure}
\begin{centering}
\includegraphics[scale=0.75]{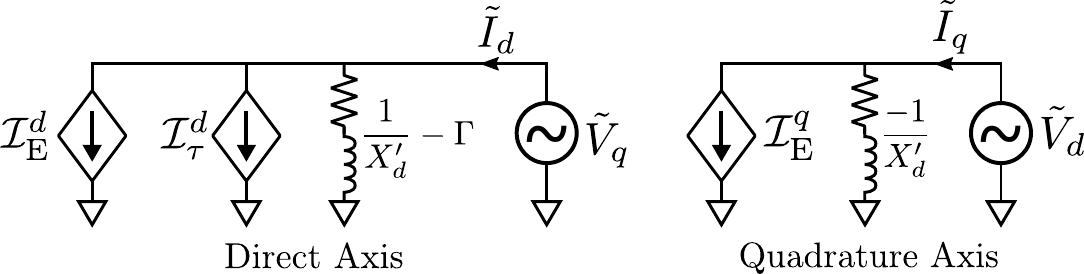} 
\par\end{centering}
\caption{\label{fig: DQ_Circuit} Circuit diagram interpretation of equation (\ref{eq: DQ_Current}), where $\mathcal{I}_{\tau}^{d}=-\Gamma\frac{X'_{d}}{{\rm E}'}\tilde{\tau}_{m}$, $\mathcal{I}_{{\rm E}}^{d}=\Gamma\frac{{\rm V}_{t}\sin(\varphi)}{{\rm E}'}\tilde{{\rm E}}'$, and $\mathcal{I}_{{\rm E}}^{q}=\frac{1}{X'_{d}}\tilde{{\rm E}}'
$ as taken from (\ref{eq: I_dqt}) and (\ref{eq: E_ph}). At non-source buses, $\mathcal{I}_{\tau}^{d}=\mathcal{I}_{{\rm E}}^{d}=\mathcal{I}_{{\rm E}}^{q}=0
$ and all current flows are caused by terminal voltage deviations.}
\end{figure}

When considering the structures of (\ref{eq: Y}), (\ref{eq: I_t}), and (\ref{eq: I_e}), it is clear that significant simplification may occur by passing to a $dq$ reference frame, i.e. rotating each expression in the direction of the rotor angle $\delta$. We use the convention of $dq$ axes orientation from~\cite{Kundur:1994}, so the rotational matrix defined as ${\rm T}_2$ is
\begin{equation}\label{eq: T2}
{\rm T}_{2}=\left[\begin{array}{cc}
\cos(\delta) & \sin(\delta)\\
-\sin(\delta) & \cos(\delta)
\end{array}\right].
\end{equation}
This transformation is applied to the state space current injection equation $\tilde{{\bf I}} =\mathcal{Y}\tilde{{\bf V}}+\mathcal{I}_{\tau}+\mathcal{I}_{{\rm E}}$ of (\ref{eq: I_Flow}). The superscript $dq$ denotes variables given in the $dq$ reference frame, while no superscript denotes variables in the real and imaginary reference frame. For instance, ${\bf{X}}=[X_r\,X_i]^{\top}$ is defined in the real and imaginary coordinate system while ${\bf{X}}^{dq}=[X_d\,X_q]^{\top}$ is defined in the $dq$ coordinate system.
\begin{align}
\tilde{{\bf I}}^{dq} & =\mathcal{Y}^{dq}\tilde{{\bf V}}^{dq}+\mathcal{I}_{\tau}^{dq}+\mathcal{I}_{{\rm E}}^{dq}\label{eq: DQ_Current},
\end{align}
where $\mathcal{Y}^{dq}={\rm T}_{2}\mathcal{Y}{\rm T}_{2}^{-1}$ and $\tilde{{\bf X}}^{dq}={\rm T}_{2}\tilde{{\bf X}}$ for any vector ${\bf X}$. Fig. \ref{fig: DQ_Qxis} provides a visualization of these transformations. In the new coordinate system, the direct ($d$) axis is in line with $\delta$, and the quadrature ($q$) axis is perpendicular to the direct axis. Once transformed, the the admittance matrix and the negative current injections are given by
\begin{align}
\mathcal{Y}^{dq} & =\left[\begin{array}{cc}
0 & \frac{1}{X'_{d}}-\Gamma\\
-\frac{1}{X'_{d}} & 0
\end{array}\right]\label{eq: Y_dq}\\
\mathcal{I}_{\tau}^{dq} & =\left[\begin{array}{c}
-\Gamma\frac{X'_{d}}{{\rm E}'}\\
0
\end{array}\right]\tilde{\tau}_{m}\label{eq: I_dqt}\\
\mathcal{I}_{{\rm E}}^{dq} & =\left[\begin{array}{c}\label{eq: E_ph}
\Gamma\frac{{\rm V}_{t}\sin\left(\varphi\right)}{{\rm E}'}\\
\frac{1}{X'_d}
\end{array}\right]\tilde{{\rm E}}'.
\end{align}
A conventional orthogonal circuit diagram interpretation of this result is given by Fig. \ref{fig: DQ_Circuit}. 
It is important to remember that $\tilde{V}_d$,  $\tilde{V}_q$, $\tilde{I}_d$, and $\tilde{I}_q$ are all complex phasors. This is a deviation from the standard power systems literature related to generator analysis (such as~\cite{Sauer:2006}) which uses orthogonal $dq$ decomposition in order to treat $V_q$ and $V_d$ as real valued signals. We note that the purpose of performing this $dq$ rotation is to build the intuition provided by equations (\ref{eq: Y_dq}-\ref{eq: E_ph}) and Fig. \ref{fig: DQ_Circuit}. In general, transforming voltages and current into a $dq$ reference frame is not necessary.

\subsection{Extension to a $6^{\rm th}$ Order Generator Model with AVR}
\begin{figure}
\begin{centering}
\includegraphics[scale=1.6]{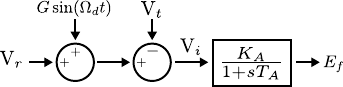} 
\par\end{centering}
\caption{\label{fig: Exciter_Model} Voltage excitation system associated with source bus \#1 in subsection \ref{sbsec: 179_FO}. The forced oscillation source is given by $G\sin(\Omega_d t)$.}
\end{figure}

Although the proposed methods for quantifying the effective admittance and current injections of a generator are developed for a low order model, the same techniques may be employed for an arbitrarily complex model. We choose to consider the source bus generator model presented in the set of standardized test cases in~\cite{Maslennikov:2016}. The source generator model may be approximated by the $6^{\rm th}$ order synchronous model presented in~\cite{Milano:2013}, where the $d$ and $q$ subscripts denote the Park reference frames. This particular model is chosen since it will be used to collect test results in Section \ref{Test_Results}:
\begin{align}
\dot{\delta} & =\Delta\omega \label{eq: delta}\\
M\Delta\dot{\omega} & =P_{{\rm m}}-P_{{\rm e}}-D\Delta\omega\\
T'_{{\rm d0}}\dot{e}'_{q} & =E_{{\rm f}}-\left(X_{d}-X'_{d}-\gamma_{d}\right)i_{d}-e'_{q}\\
T'_{{\rm q0}}\dot{e}'_{d} & =\left(X_{q}-X'_{q}-\gamma_{q}\right)i_{q}-e'_{d}\\
T''_{d0}\dot{e}''_{q} & =e'_{q}-e''_{q}-\left(X'_{d}-X''_{d}+\gamma_{d}\right)i_{d}\\
T''_{q0}\dot{e}''_{d} & =e'_{d}-e''_{d}+\left(X'_{q}-X''_{q}+\gamma_{q}\right)i_{q}\label{eq: edpp},
\end{align}
where $\gamma_{x}=T''_{x0}X''_{x}\left(X_{x}-X'_{x}\right)/\left(T'_{x0}X'_{x}\right)$, $x\in\{d,q\}$. With stator resistance neglected, the electrical power is $P_{{\rm e}}=e_d i_d + e_q i_q$, and the terminal currents ($i_d$, $i_q$) can be written in terms of the terminal voltages ($e_{d}={\rm V}_{t}\sin\left(\delta-\theta_{t}\right)$,  $e_{q}={\rm V}_{t}\cos\left(\delta-\theta_{t}\right)$) and the subtransient voltages ($e''_d$, $e''_q$):
\begin{equation}\label{eq: idiq}
\left[\begin{array}{c}
i_{d}\\
i_{q}
\end{array}\right]=\left[\begin{array}{cc}
R & -X_{q}''\\
X_{d}'' & R
\end{array}\right]^{-1}\left[\begin{array}{c}
e''_{d}-{\rm V}_{t}\sin\left(\delta-\theta_{t}\right)\\
e''_{q}-{\rm V}_{t}\cos\left(\delta-\theta_{t}\right)
\end{array}\right],
\end{equation}
where $R=0$ when neglected. The real and imaginary \textbf{negative} current injections are computed by simply rotating $i_{d}$ and $i_{q}$ in rectangular space~\cite{Sauer:2006} and negating. Equation (\ref{eq: iRiI}) is a time domain transformation and should not be confused with the phasor reference frame transformation of (\ref{eq: T2}):
\begin{align}\label{eq: iRiI}
\left[\begin{array}{c}I_R\\
I_I
\end{array}\right]=-\left[\begin{array}{cc}
\sin\left(\delta\right) & \cos\left(\delta\right)\\
-\cos\left(\delta\right) & \sin\left(\delta\right)
\end{array}\right]\left[\begin{array}{c}
i_{d}\\
i_{q}
\end{array}\right].
\end{align}
Finally, since PMUs measure the magnitude and phase of voltage and current signals, it is numerically convenient to have the generator's FRF relate voltage magnitude and phase perturbations with current magnitude and phase perturbations. Therefore, the generator model needs some nonlinear function relating its state and algebraic variables to the current magnitude ($\rm I$) and current phase ($\phi$):
\begin{align}
{\rm I} & =\sqrt{{I_R}^{2}+{I_I}^{2}}\\
\phi & =\tan^{-1}\left(\frac{I_I}{I_R}\right).
\end{align}Controllers may also be included in the generator model. The static voltage excitation system associated with the source generator of test case ``F1" in~\cite{Maslennikov:2016} is approximated by the block diagram in Fig. \ref{fig: Exciter_Model} (limits excluded). The source of the forced oscillation is given by $G\sin(\Omega_d t)$ with gain $G$ and forcing frequency $\Omega_d$. This forcing function is not included in the system model; it is only shown for illustration.
The exciter's associated differential equation follows:
\begin{equation}\label{eq: Exc}
T_{A}\dot{E}_{f}=K_{A}{\rm V}_{i}-E_{f}.
\end{equation}
Now that the generator's full set of nonlinear Differential Algebraic Equations (DAEs), $f$ and $g$ respectively, have been specified, they can be written as follows, with state variable vector $\mathbf{x}$, algebraic variable vector $\mathbf{y}=[{\rm V} \; \theta]^\top$, and output vector ${\bf I}=[{\rm I} \; \phi]^\top$:
\begin{align}\label{eq: f}
\dot{\mathbf{x}} & =f\left(\mathbf{x},\mathbf{y}\right)\\
{\bf I} & =g\left(\mathbf{x},\mathbf{y}\right)\label{eq: g}.
\end{align}
These DAEs are linearized such that $\Delta\dot{\mathbf{x}} =f_{{\bf x}}\Delta\mathbf{x}+f_{{\bf y}}\Delta\mathbf{y}$ and $\Delta{\bf I}=g_{{\bf x}}\Delta\mathbf{x}+g_{{\bf y}}\Delta\mathbf{y}$. Finally, the generator's FRF ${\mathcal Y}$ can be built:
\begin{align}\label{eq: FRF}
\text{FRF} \rightarrow \;\; \mathcal{Y}=g_{\bf x}(j\Omega{\mathbb 1}-f_{\bf x})^{-1}f_{\bf y}+g_{\bf y}.
\end{align}
This FRF relates the Fourier transform of the inputs and the outputs across the full spectrum frequencies. In defining the Fourier transform of the time domain signal $x(t)$ as $\tilde x(\Omega)= \int_{-\infty}^{+\infty}x(t)e^{-j\Omega t}{\rm d}t$, we see that the the FRF relates the Fourier transforms of the time domain voltages and currents in the following way:
\begin{align}\label{eq: Admittance_Test}
\left[\begin{array}{c}
\tilde{{\rm I}}(\Omega)\\
\tilde{\phi}(\Omega)
\end{array}\right]=\mathcal{Y}(\Omega)\left[\begin{array}{c}
\tilde{{\rm V}}(\Omega)\\
\tilde{\theta}(\Omega)
\end{array}\right],\quad\Omega\in[0\;\;\infty).
\end{align}
Of course, generators are complex machines which may have a variety of controllers (AVR, PSS, etc.) and a multitude of states, but this process may be generalized for arbitrarily complex DAE systems $f$ and $g$ so long as terminal current can be written as a function of terminal voltage.


\section{Leveraging $\mathcal{Y}$ For Source Detection}\label{Method}
In a deterministic power system where generator model parameters are fully known, measurement noise is negligible and perturbations are small, the FRF ${\mathcal Y}$ can fully predict the measured spectrum of the generator output ${\tilde {\bf I}}$ for a given measured spectrum input ${\tilde {\bf V}}$ at all non-source generators. In this \textit{ideal} system, the following simple test may be naively applied at each generator across the full spectrum of frequencies.
\begin{align}
\tilde{{\bf I}} & =\mathcal{Y}\tilde{{\bf V}}\;\rightarrow\;\text{Non-source generator}\label{eq: Test_1}\\
\tilde{{\bf I}} & \ne\mathcal{Y}\tilde{{\bf V}}\;\rightarrow\;\text{Source generator}\label{eq: Test_2}
\end{align}
In other words, if the measured current spectrum ${\bf {\tilde I}}$ and the predicted current spectrum $\mathcal{Y}{\bf {\tilde V}}$ match, then the generator has no internal oscillation source. If, though, ${\bf {\tilde I}} \ne \mathcal{Y}{\bf {\tilde V}}$ at some particular frequency, then a current source (forced oscillation) may be present in the generator at said frequency. Of course, to implement this test on any given generator, there must be a PMU present which is capable of measuring the generator's terminal voltages and currents so that their respective spectrums may be computed.

The realities of power system operation can prevent the naive tests of (\ref{eq: Test_1}) and (\ref{eq: Test_2}) from being directly implemented. There are three primary sources of potential error in this process. First, nonlinearities may prevent the admittance matrix, which is built on a linearized system model, from exactly predicting the generator dynamics. The extent of nonlinear system behavior depends on the size of the oscillation, but the associated error is typically small enough to be neglected. Secondly, in building the FRF, generator parameters (damping, time constants, etc.) may have a large degree of uncertainty. Accordingly, the results presented from tests on the 179-bus system in section \ref{Test_Results} consider this uncertainty. And thirdly, despite the fact that IEEE Standard C37.242 specifics that PMU magnitude error must be below 0.1\%, and timing error must be better than 1 $\mu$s (or $0.02\degree$)~\cite{PMU_Std}, additive error from current and voltage transformer equipment may present additional error. Since measured voltage and current spectral comparisons can breakdown severely when this nontrivial PMU measurement noise is present, the next section introduces a framework for dealing with the problem of additive measurement noise.

\subsection{Bounding Error Associated with PMU Measurement Noise}
We define ${\rm V}(t)$, $\theta(t)$, ${\rm I}(t)$ and ${\phi}(t)$ to be the true voltage magnitude, voltage phase, current magnitude, and current phase time series vectors, respectively, at some generator bus. We further assume these vectors are perturbations from their respective steady state operating points. We now define the \textit{measured} time series vectors to be $\hat{X}(t)$, where the true signals are corrupted by Additive White Gaussian Noise (AWGN) from $\epsilon_X(t)$:
\begin{align}
\hat{X}(t) & =X(t)+\epsilon_{X}(t),\;\;\;\;\;\;X\in[{\rm V},\;\theta,\;{\rm I},\;\phi].
\end{align}
In measuring the spectrum of $\hat{X}(t)$, we invoke the linearity property of the Fourier transform, such that
\begin{align}
\mathcal{F}\{\hat{X}(t)\} & =\tilde{\hat{X}}(\Omega)\\
 & \coloneqq \tilde{X}(\Omega)+\tilde{\epsilon}_{X}(\Omega).
\end{align}
The Fourier transform of AWGN will ideally have a flat magnitude spectrum (equal to $\lambda_{\epsilon_{X}}$) and a uniformly distributed phase spectrum characterized by $\mathcal{U}(0,2\pi)$:
\begin{equation}
\tilde{\epsilon}_{X}(\Omega)=\lambda_{\epsilon_{X}}e^{j\mathcal{U}(0,2\pi)},\;\;\;\;\Omega\in[0\;\;\infty)\label{eq: PMU_noise_dist}.
\end{equation}
In applying the admittance matrix transformation of (\ref{eq: Admittance_Test}) to calculate the difference in the measured (${\bf \tilde{I}}$) and the predicted ($\mathcal{Y}{\bf \tilde{V}}$) currents at some non-source bus, the following error may be approximated:
{\small{}\begin{align}
{\bf \tilde{I}}-\mathcal{Y}{\bf \tilde{V}} & =\left[\begin{array}{c}
(\tilde{{\rm I}}+\tilde{\epsilon}_{{\rm I}})-\mathcal{Y}_{11}(\tilde{{\rm V}}+\tilde{\epsilon}_{{\rm V}})-\mathcal{Y}_{12}(\tilde{\theta}+\tilde{\epsilon}_{\theta})\\
(\tilde{\phi}+\tilde{\epsilon}_{\phi})-\mathcal{Y}_{21}(\tilde{{\rm V}}+\tilde{\epsilon}_{{\rm V}})-\mathcal{Y}_{22}(\tilde{\theta}+\tilde{\epsilon}_{\theta})
\end{array}\right]\\
 & \approx\left[\begin{array}{c}\label{eq: diff_approx}
\tilde{\epsilon}_{{\rm I}}-\mathcal{Y}_{11}\tilde{\epsilon}_{{\rm V}}-\mathcal{Y}_{12}\tilde{\epsilon}_{\theta}\\
\tilde{\epsilon}_{\phi}-\mathcal{Y}_{21}\tilde{\epsilon}_{{\rm V}}-\mathcal{Y}_{22}\tilde{\epsilon}_{\theta}
\end{array}\right]\\
 & \coloneqq\left[\begin{array}{c}\label{eq: error_vec}
\tilde{\epsilon}_{\bf m}\\
\tilde{\epsilon}_{\bf p}
\end{array}\right],
\end{align}}{\small}where the simplification in (\ref{eq: diff_approx}) is due to the fact that, theoretically, $\tilde{{\rm I}}-\mathcal{Y}_{11}\tilde{{\rm V}}-\mathcal{Y}_{12}\tilde{\theta}=0$ and $\tilde{\phi}-\mathcal{Y}_{21}\tilde{{\rm V}}-\mathcal{Y}_{22}\tilde{\theta}=0$ for all frequencies. In (\ref{eq: error_vec}), the variables $\tilde{\epsilon}_{\bf m}$ and $\tilde{\epsilon}_{\bf p}$ have been defined which represent the aggregate measurement error spectrums associated with ${\bf \tilde{I}}-\mathcal{Y}{\bf \tilde{V}}$. We seek to quantify this error, at each frequency $\Omega$, with the $l2$ norm such that
\begin{align}
\left\Vert {\bf \tilde{I}}-\mathcal{Y}{\bf \tilde{V}}\right\Vert _{2}=\sqrt{\left|\tilde{\epsilon}_{{\bf m}}\right|^{2}+\left|\tilde{\epsilon}_{{\bf p}}\right|^{2}}.
\end{align}
As can been seen from (\ref{eq: diff_approx}), this error norm will be maximized when the complex entries meet the following phase conditions: 
\begin{align}
\angle\tilde{\epsilon}_{{\rm I}} &= -\angle(\mathcal{Y}_{11}\tilde{\epsilon}_{{\rm V}}) = -\angle(\mathcal{Y}_{12}\tilde{\epsilon}_{\theta})\\
\angle\tilde{\epsilon}_{\phi} &= -\angle(\mathcal{Y}_{21}\tilde{\epsilon}_{{\rm V}}) = -\angle(\mathcal{Y}_{22}\tilde{\epsilon}_{\theta}).
\end{align}
Since the measurement error spectrums have uniformly distributed phase angles $\mathcal{U}(0,2\pi)$, this is a plausible scenario and it provides us with a theoretical upper bound on the measurement error for a generator with known model parameters and no forced oscillation:
\begin{align}\label{eq: max_l2error}
\Sigma_{2} & \coloneqq \sqrt{{\rm max}\left\{ \tilde{\epsilon}_{{\bf m}}\right\}^{2}+{\rm max}\left\{ \tilde{\epsilon}_{{\bf p}}\right\}^{2}},
\end{align}
where we give the following definitions for ${\rm max}\left\{ \tilde{\epsilon}_{{\bf m}}\right\}$ and ${\rm max}\left\{ \tilde{\epsilon}_{{\bf p}}\right\}$:
\begin{align}
{\rm max}\left\{ \tilde{\epsilon}_{{\bf m}}\right\}\label{eq: em}  & =\left|\tilde{\epsilon}_{{\rm I}}\right|+\left|\mathcal{Y}_{11}\tilde{\epsilon}_{{\rm V}}\right|+\left|\mathcal{Y}_{12}\tilde{\epsilon}_{\theta}\right|\\
{\rm max}\left\{ \tilde{\epsilon}_{{\bf p}}\right\}\label{eq: ep}  & =\left|\tilde{\epsilon}_{\phi}\right|+\left|\mathcal{Y}_{21}\tilde{\epsilon}_{{\rm V}}\right|+\left|\mathcal{Y}_{22}\tilde{\epsilon}_{\theta}\right|.
\end{align}
In (\ref{eq: max_l2error}), $\Sigma_{2}$ is the maximum upper bound on the aggregate measurement error, and it is uniquely defined for all frequencies since both $\tilde{\epsilon}_{{\bf m}}$ and $\tilde{\epsilon}_{{\bf p}}$ are direct functions of frequency. If $\Vert {\bf \tilde{I}}-\mathcal{Y}{\bf \tilde{V}}\Vert _{2}$ is significantly larger than $\Sigma_{2}$ at some frequency, then PMU measurement error may not be the source of the error, and an internal forced oscillation may be to blame. In calculating (\ref{eq: em}) and (\ref{eq: ep}), the operator must have a sense of the PMU measurement noise strength. Ideally, this noise strength is constant in the frequency domain, but realistically, it fluctuates for a time limited signal $\epsilon_{X}(t)$. Therefore, in estimating the measurement noise strength in any PMU signal, a system operator should be conservative in choosing values for $\lambda_{\epsilon_{X}}$ from (\ref{eq: PMU_noise_dist}). One such conservative choice, which has been found via experimentation, is to set $\lambda_{\epsilon_{X}}$ equal to twice the expected value of the magnitude of the fast Fourier transform (fft) of its associated time domain signal $\epsilon_{X}(t)$, where $\epsilon_{X}(t)$ is constructed by sampling length($t$) times from $\mathcal{N}(0,\sigma^2_{\rm PMU})$. Therefore,
\begin{equation}\label{eq: assume_lambda}
\lambda_{\epsilon_{X}}\approx 2 \cdot {\rm E}\left[ \left|{\rm fft}\left\{ \epsilon_{X}(t)\right\} \right|\right].
\end{equation}

\subsection{Defining a Practical Source Location Technique}
In computing the error between the measured and predicted currents at a given bus, (\ref{eq: max_l2error}) defines a useful approximate upper bound on the associated measurement error. As long as the strength of the measurement noise is known (or can be estimated, such as in~\cite{Brown:2016}), this upper bound can be computed for all frequencies. Assuming an accurate FRF, significant deviations from this upper bound at any given frequency may indicate the presence of an internal current source (forced oscillation). To quantify the size of the spectral deviation at each frequency, we introduce a metric termed the Local Spectral Deviation (LSD). Its form is given as follows:
\begin{align}\label{eq: LSD}
{\rm LSD}= & \left\Vert {\bf \tilde{I}}-\mathcal{Y}{\bf \tilde{V}}\right\Vert _{2}-\Sigma_{2}.
\end{align}
Table \ref{tab: LSD_Terms} summarizes the terms in (\ref{eq: LSD}) which is computed at all generators for which terminal PMU data data is available. Formally, the LSD calculates the difference in the prediction error and the maximum bound on the effects of measurement noise error. To apply the LSD, the operator should first determine the central forcing frequency $\Omega_d$ of the system (there may be multiple forcing frequencies if the system is experiencing multiple forced oscillations).
\begin{table}
\begin{centering}
\caption{\label{tab: LSD_Terms}Definition of LSD Terms from (\ref{eq: LSD})}
\begin{tabular}{cl}
\hline 
\noalign{\vskip\doublerulesep}
$\tilde{{\bf I}}$ & {\bf Measured} $2\times1$ vector of complex valued current magnitude\tabularnewline
 & and phase variables ${\tilde {\rm I}}(\Omega)$ and ${\tilde {\phi}}(\Omega)$ \tabularnewline
$\tilde{{\bf V}}$ & {\bf Measured} $2\times1$ vector of complex valued voltage magnitude\tabularnewline
 & and phase variables ${\tilde {\rm V}}(\Omega)$ and ${\tilde {\theta}}(\Omega)$ \tabularnewline
${\mathcal Y}   $ & {\bf Modeled} $2\times2$ frequency dependent complex admittance\tabularnewline
& matrix, as given by (\ref{eq: FRF}) \tabularnewline
$\Sigma_2$ & {\bf Estimated} upper bound (frequency dependent) on measurement error\tabularnewline
& effects, as given by the maximum $l2$ norm of the vector in (\ref{eq: error_vec})
\vspace{2mm}\tabularnewline
\hline 
\end{tabular}
\par\end{centering}
\end{table}
In Algorithm (\ref{alg: FO}), the steps for using generator terminal data to determine whether or not a generator is the source of a forced oscillation are formalized. In this algorithm, the operator specified threshold $\iota$ is used to determine if the LSD is large enough for a generator to be deemed a source.

We note that this algorithm should be applied in situations where an operator has a high degree of certainty that the detected oscillations are in fact forced oscillations (references such as~\cite{Xie:2015} and~\cite{Ye:2017} can be useful to this end); the method we have developed will not locate the source of negative damping in a system, and therefore it will be unhelpful in locating the source of a natural oscillation.

\begin{algorithm}\label{alg: FO}
\caption{Generator Source Detection Method}
\textbf{START}\\
\begin{enumerate}[label=\textbf{\arabic*},start=1]
	\item Use available generator model data to construct\\
           DEA sets (\ref{eq: f}) and (\ref{eq: g}).
    \item Build the FRF $\mathcal{Y}$ of (\ref{eq: FRF}) which relates polar\\
          voltage and polar current deviations
    \item Acquire PMU time series vectors ${\rm V}(t)$, $\theta(t)$, ${\rm I}(t)$,\\
          and $\phi(t)$ from the generator terminals
    \item Subtract estimated steady state operating points\\
          from these time series vectors
    \item Take the fft of these perturbation vectors to\\
          build ${\tilde {\bf I}}(\Omega)$ and ${\tilde {\bf V}}(\Omega)$
    \item Identify forcing frequency (or frequencies) $\Omega_d$
    \item Compute the LSD of (\ref{eq: LSD}) at $\Omega_d$
\end{enumerate}
 \uIf{${\rm LSD}<0$}
  {Prediction error is less than $\Sigma_2$: \\
  \begin{itemize}
   \item Generator \textbf{is not} a source
  \end{itemize}}
  \uElseIf{$0<{\rm LSD}<\iota$}{Prediction error is larger than $\Sigma_2$ but less that $\iota$:
                                \begin{itemize}
                                \item Generator \textbf{probably not} a source
                                \end{itemize}}
  \Else{Prediction error is larger than threshold:
                                \begin{itemize}
                                \item Generator \textbf{is} a source
                                \end{itemize}}
\end{algorithm}
\vspace*{-.5cm}


\section{Test Results}\label{Test_Results}
In this section, we present five sets of test results. First, we consider a 3-bus system of two $2^{\rm nd}$ order generators tied to an infinite bus. Second, we test our method on the modified WECC 179-bus system in the presence of a forced oscillation. Third, we test our method on the modified WECC 179-bus system in the presence of a natural oscillation. Fourth, we apply a rectangular forced oscillation in the WECC 179-bus system when a poorly damped mode is present. And fifth, we contrast the effectiveness of the DEF method and the FRF source location method in the context of a three-bus system with a constant impedance load.
\subsection{Radial Generators Tied to Infinite Bus}
\begin{figure}
\begin{centering}
\includegraphics[scale=0.9]{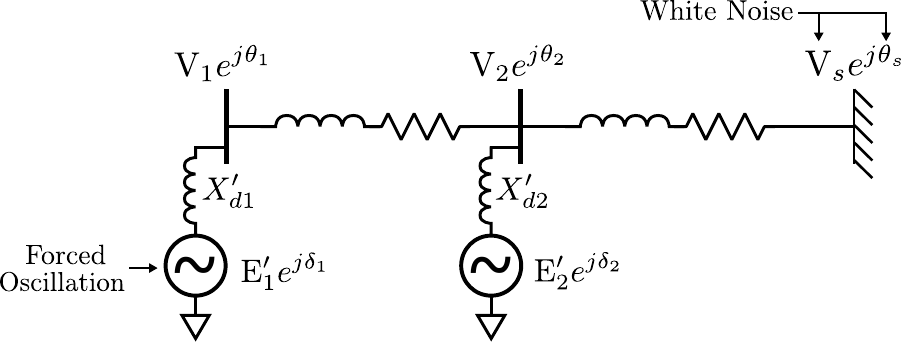} 
\par\end{centering}
\caption{\label{fig: 3_Bus_Inf_-_Rev} 3 Bus Diagram with Infinite Bus. Both generators are $2^{\rm nd}$ order, and a mechanically forced oscillation $\tau_m=\tau_0 + {\tilde \tau}$ is placed on generator 1. White noise is applied to the phase and magnitude of the infinite bus voltage.}
\end{figure}

It is well known in the literature \cite{OBrien:2017,Wilson:2014} that relying on the location of the largest detected oscillations is an unreliable way for determining the source of a forced oscillation. Because of the excitation of local resonances, large power oscillations can occur at non-source generators. We demonstrate the effectiveness of our source location technique in the presence of resonance amplification occurring on a non-source generator by simulating the simple 3-bus system of two radial generators tied to an infinite bus as given by Fig. \ref{fig: 3_Bus_Inf_-_Rev}. In this system, the lines have $X=0.1$ and $R=0.01$, and other system parameters are summarized in Table \ref{tab: Gen_Data}. A forced oscillation is applied to the mechanical torque of generator 1 via $\tau_m=\tau_0 + \alpha\sin(\Omega_dt)$. Additionally, ambient white noise is applied to the magnitude and the phase of the infinite bus voltage to mimic system fluctuations.

\begin{figure}
\begin{centering}
\includegraphics[scale=0.44]{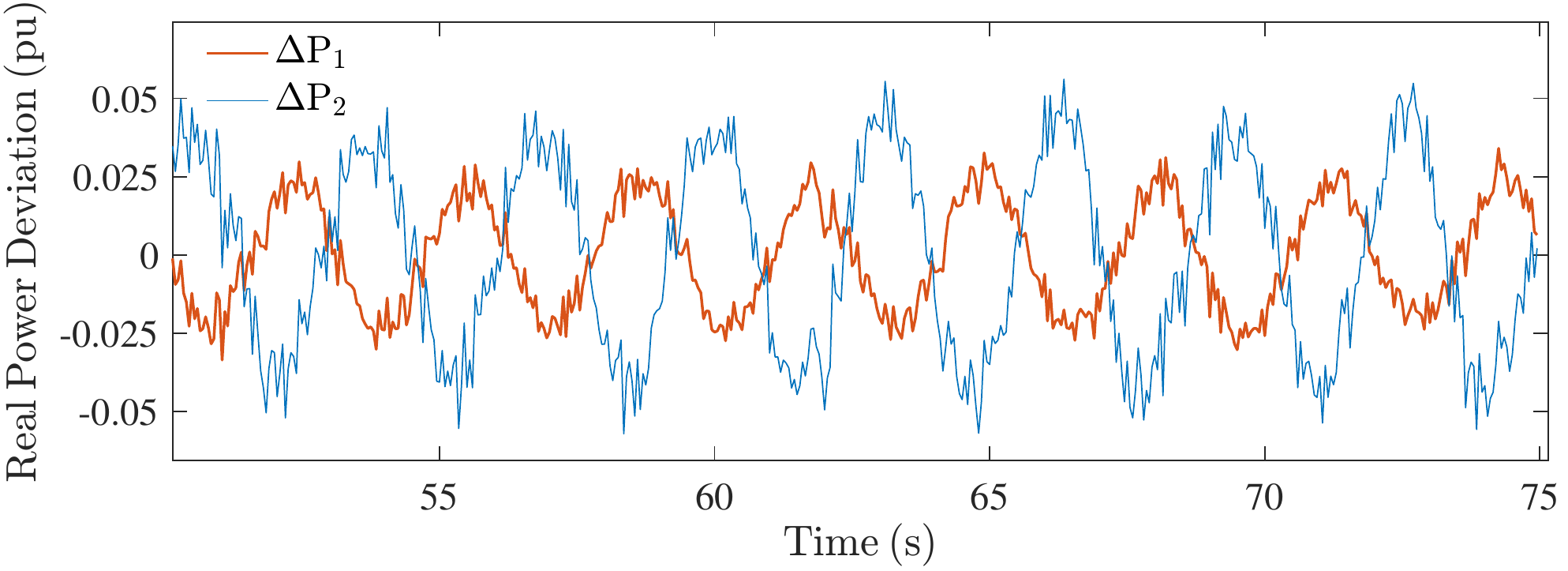} 
\par\end{centering}
\caption{\label{fig: Active_Power_Injection} Active power injection deviations for generators 1 and 2.}
\end{figure}

The driving frequency of the forced oscillation $\Omega_d$ is chosen by considering the eigenvalues of the system. To find these eigenvalues, the system DAEs of $\dot{\mathbf{x}} =f\left(\mathbf{x},\mathbf{y}\right)$ and $0 =g\left(\mathbf{x},\mathbf{y}\right)$ were linearized such that $\Delta\dot{\mathbf{x}} =f_{{\bf x}}\Delta\mathbf{x}+f_{{\bf y}}\Delta\mathbf{y}$ and $0=g_{{\bf x}}\Delta\mathbf{x}+g_{{\bf y}}\Delta\mathbf{y}$. The imaginary parts of the complex eigenvalues of the state matrix $A_s=f_{{\bf x}}-f_{{\bf y}}g_{{\bf y}}^{-1}g_{{\bf x}}$ yield the set of natural frequencies. The natural modes associated with generators 1 and 2 are $\Omega_{d1}=0.708$ $\frac{{\rm rad}}{{\rm sec}}$ and $\Omega_{d2}=1.915$ $\frac{{\rm rad}}{{\rm sec}}$, respectively. We therefore choose to mechanically force the system at $f_d=2$ since this is close to, but not directly on top of, the natural mode of generator 2. Fig. \ref{fig: Active_Power_Injection} shows a time domain simulation plot of the power injection deviations at each generator. The standard deviation of power injections at generator 2 is almost twice as larger as that of generator 1, and the forcing frequency of $f_d=\frac{2}{2\pi}$ Hz can be seen underneath the system noise.

\begin{figure}
\begin{centering}
\includegraphics[scale=0.44]{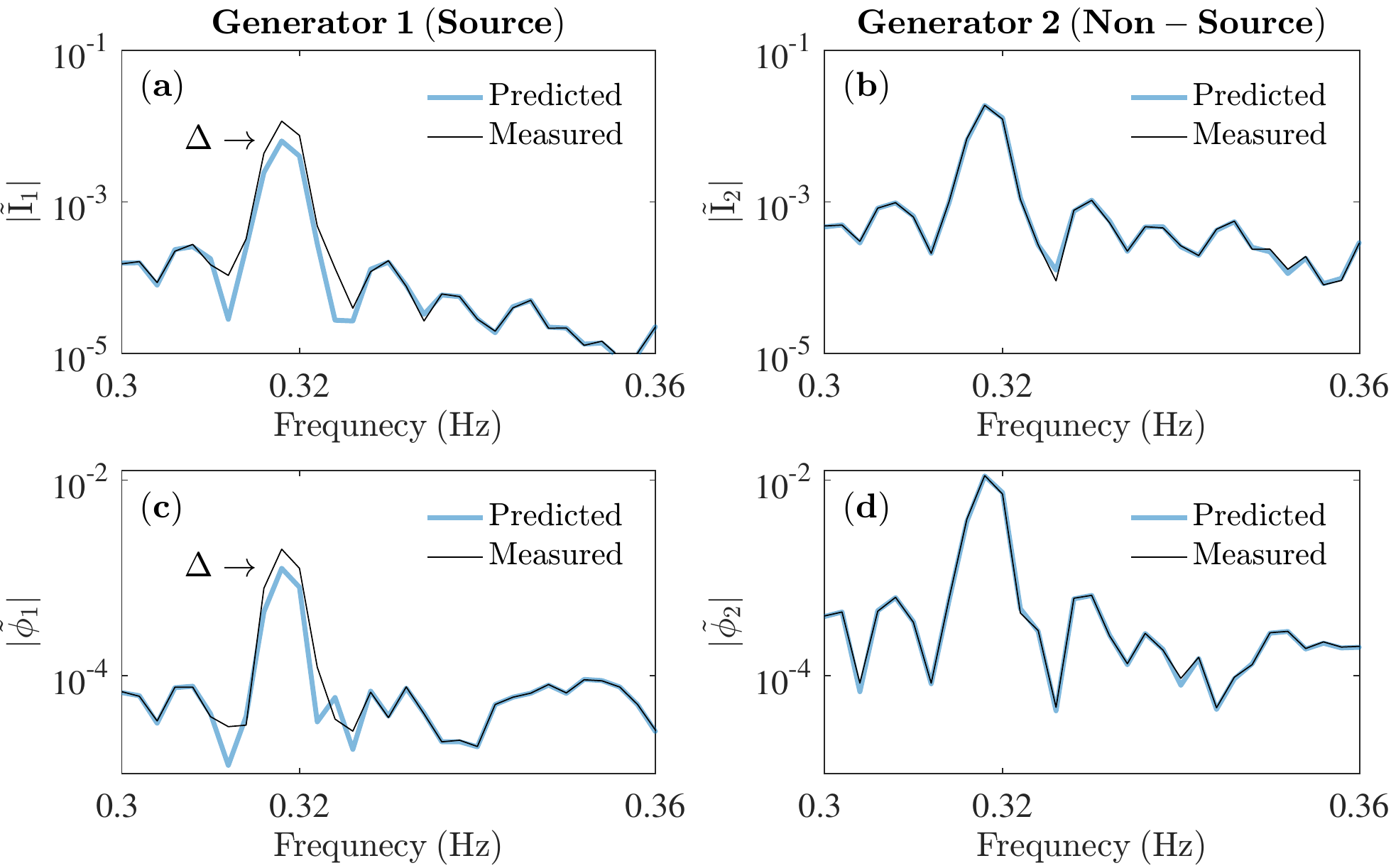} 
\par\end{centering}
\caption{\label{fig: 3_Bus_PSD} Spectral magnitude of current magnitude (panels $({\bf a})$ \& $({\bf b})$) and current phase (panels $({\bf c})$ \& $({\bf d})$) perturbations are given for each generator. The forcing frequency is located at $f_d=0.32$. The $\Delta$ symbol in panels $({\bf a})$ and $({\bf c})$ highlight the locations of significant deviations between the predicted and expected spectrums.}
\end{figure}

After collecting the time domain voltage and current data from the simulation, the predicted ($\mathcal{Y}{\tilde {\bf V}}$) and measured (${\tilde {\bf I}}$) current spectrums were compared. For illustrative purposes, measurement noise is not applied and generator model parameter uncertainty is neglected such that ${\mathcal Y}$ is known exactly for both generators. For a small frequency range, the magnitude spectrum comparisons are given by Fig. \ref{fig: 3_Bus_PSD}. There are two important observations concerning these comparisons. First, the spectral peaks of generator 2 (the non-source generator) at the forcing frequency of $f_d=0.32$ are much larger than the spectral peaks of generator 1 (the source generator) due to resonance. Second, the predicted and measured spectrums at the forcing frequency of the source generator (seen in panels $({\bf a})$ \& $({\bf c})$) misalign significantly. From direct visual inspection of Fig. \ref{fig: 3_Bus_PSD}, it is clear that a modest internal oscillation is present on generator 1 which is causing deviations between the measured and the predicted spectrums (the LSD is not computed since measurement noise is not applied in this test).

\begin{table}
\caption{\label{tab: Gen_Data} Generator Parameters}
\centering{}
\begin{tabular}{|c|c|c|c|c|c|c|}
\hline 
 & $M$ & $D$ & $X'_{d}$ & ${\rm E}'$ & ${\rm V}_{t}$ & $\varphi$\tabularnewline
\hline 
\hline 
Gen 1 & $4$ & $1$ & $0.25$ & $1.019$ & $1$ & $0.248$\tabularnewline
\hline 
Gen 2 & $1$ & $0.25$ & $0.2$ & $1.031$ & $1$ & $0.117$\tabularnewline
\hline 
\end{tabular}
\end{table}

\subsection{WECC 179-Bus System (Forced Oscillation)}\label{sbsec: 179_FO}
\begin{figure}
\begin{centering}
\includegraphics[scale=0.45]{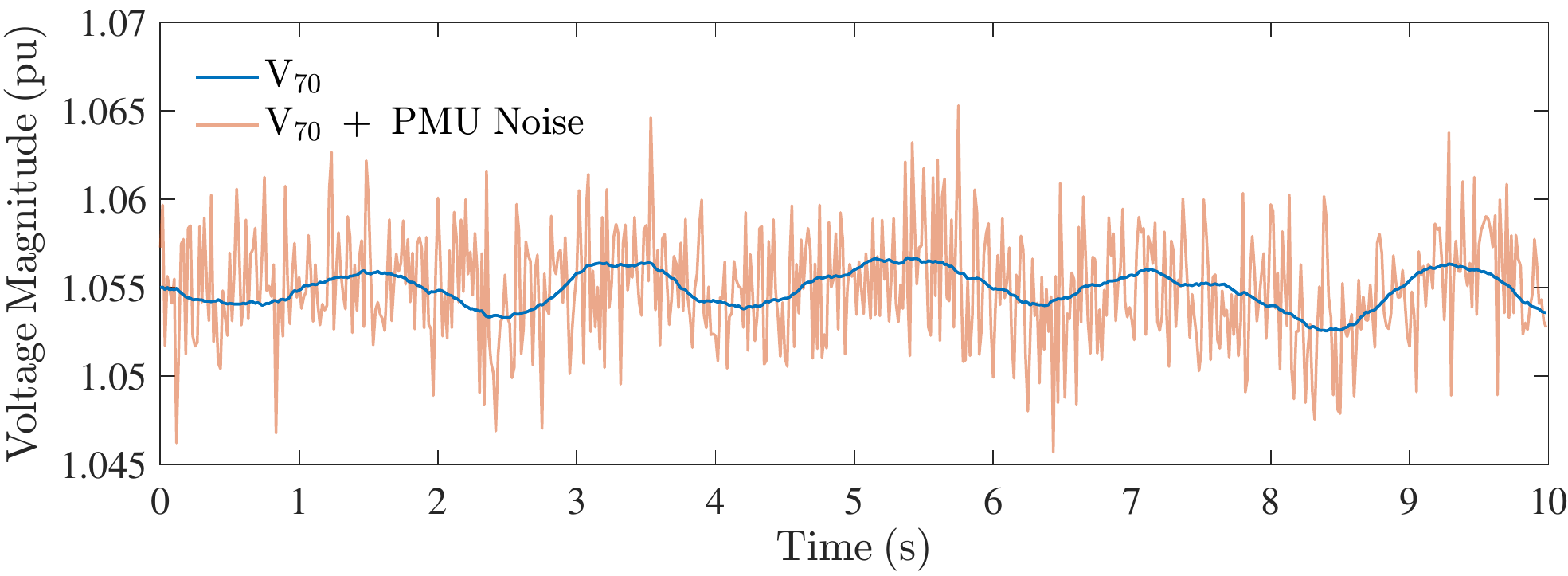} 
\par\end{centering}
\caption{\label{fig: V70_PMU} Actual and measured voltage magnitude at bus 70 (generator).}
\end{figure}

For further validation, we apply these methods on data collected from the WECC 179-bus system in the presence of multiple forced oscillations. As suggested in~\cite{Maslennikov:2016}, the standardized test case files were downloaded and simulated using Power Systems Analysis Toolbox (PSAT)~\cite{Milano:2013}. We chose to investigate the performance of our methods on a modified version of test case ``F1". In ``F1", a scaled 0.86 Hz sinusoid is added to the reference signal of the AVR attached to the source generator at bus 4 (see~\cite{Maslennikov:2016} for a full system map). In the system, all loads are constant power while all non-source generators are modeled as $2^{\rm nd}$ order classical machines with parameters $D=4$, $X'_d=0.25$, and various inertias around $H=3$ (machine base). The source generator is a sixth order synchronous machine with an Automatic Voltage Regulator (AVR) modeled by Fig. \ref{fig: Exciter_Model}. To engender a realistic testing scenario, we modify this test case in three major ways. 
\begin{enumerate}
\item
Load fluctuations are added to all PQ loads. The dynamics of these fluctuations are modeled by the Ornstein-Uhlenbeck process~\cite{Ghanavati:2016} of
\begin{align}
\mathbf{\dot{u}}(t)&=-E\mathbf{u}(t)+{\mathbb 1}\underline{\xi}\label{eq: u_dot},
\end{align}
where ${\mathbb 1}$ is the $n$x$n$ identity matrix for $n$ PQ loads and $E$ is a diagonal matrix of inverse time correlations. $\underline{\xi}$ is a vector of zero-mean independent Gaussian random variables (standard deviation $\sigma=2.5e-3$). The noise vector ${\bf u}(t)$ is added to the PQ loads such that
\begin{align}
{\bf S}(t)&={\bf S}_{0}\left(1+{\bf u}(t)\right)
\end{align}
where ${\bf S}(t)={\bf P}(t)+j{\bf Q}(t)$.
\item
Two additional forced oscillations are added to the system (along with the AVR oscillation at generator bus 4). Each new oscillation is added to the mechanical torque of a $2^{\rm nd}$ order system generator according to
\begin{equation}
\tau_m=\tau_0(1+\alpha_i\sin(\Omega_{d_i}t)).
\end{equation}
These forced oscillations are arbitrarily added to generator buses 13 and 65, and in each case $\alpha_i=0.05$. One of these oscillations is applied at $f_d=0.5$ Hz and the second is applied at $f_d=2.0$ Hz.
\item
PMU measurement noise is added to the simulation data. AWGN with a standard deviation of $\sigma=0.3\;(\% \; {\rm pu})$ is applied to all PMU times series vectors. This value of $\sigma$ was chosen since the associated distribution tails realistically extend up to $\pm$1\% pu. For a visualization of the effect of PMU measurement noise in the presence of system dynamics, Fig. \ref{fig: V70_PMU} shows the bus voltage magnitude of a generator bus (bus 70). The applied noise greatly corrupts the fft calculations.
\end{enumerate}

\begin{figure}
\includegraphics[scale=0.45]{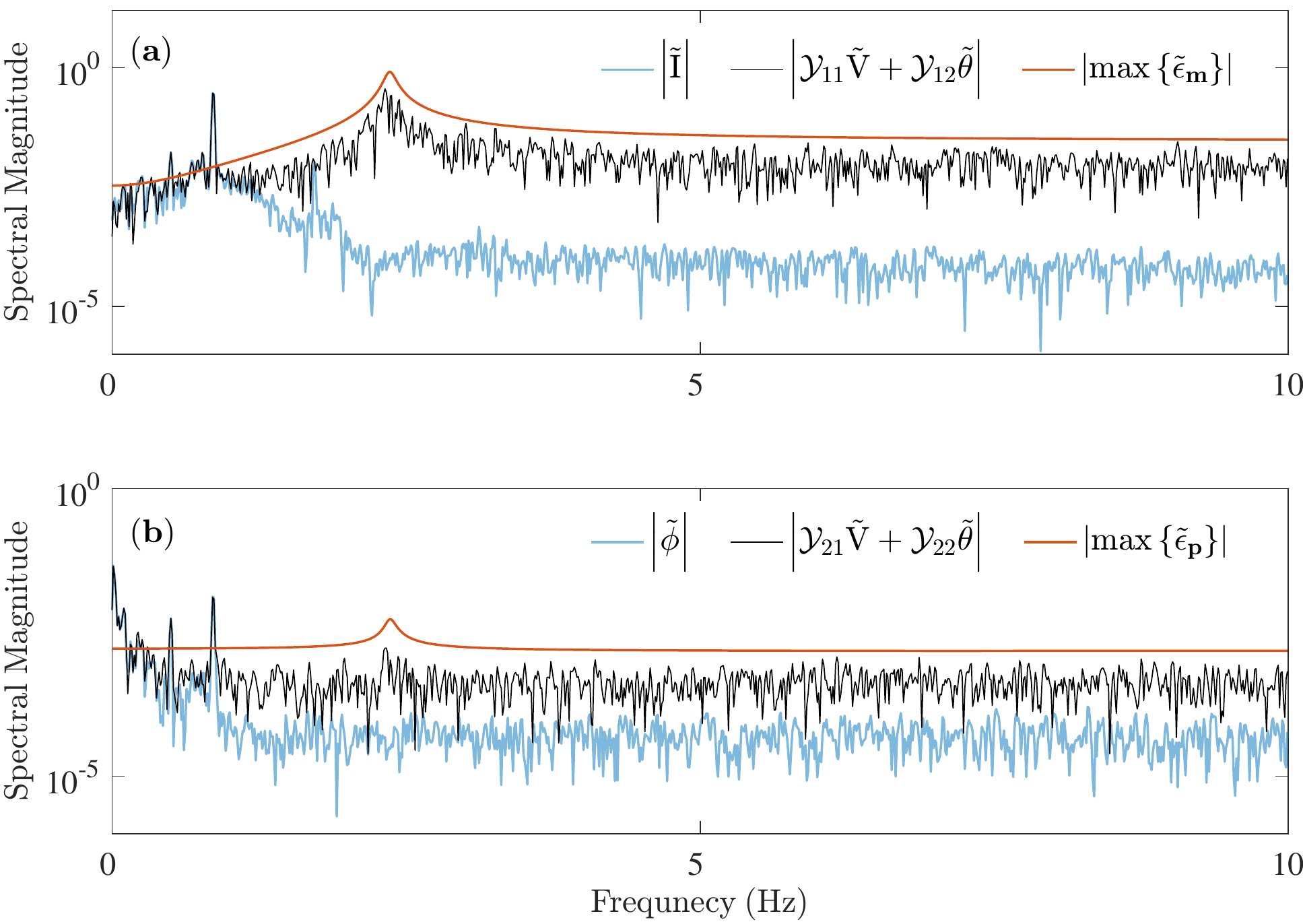}
\caption{\label{fig: Mag_Gen3} The spectral magnitude of the measured current magnitude (panel $({\bf a})$) and the measured current phase (panel $({\bf b})$) perturbations at generator bus 9 are given by the blue traces. The associated predicted spectral magnitudes are given by the black traces. Finally, the orange traces give the estimated maximum PMU measurement noise errors.}
\end{figure}

\begin{figure}
\includegraphics[scale=0.45]{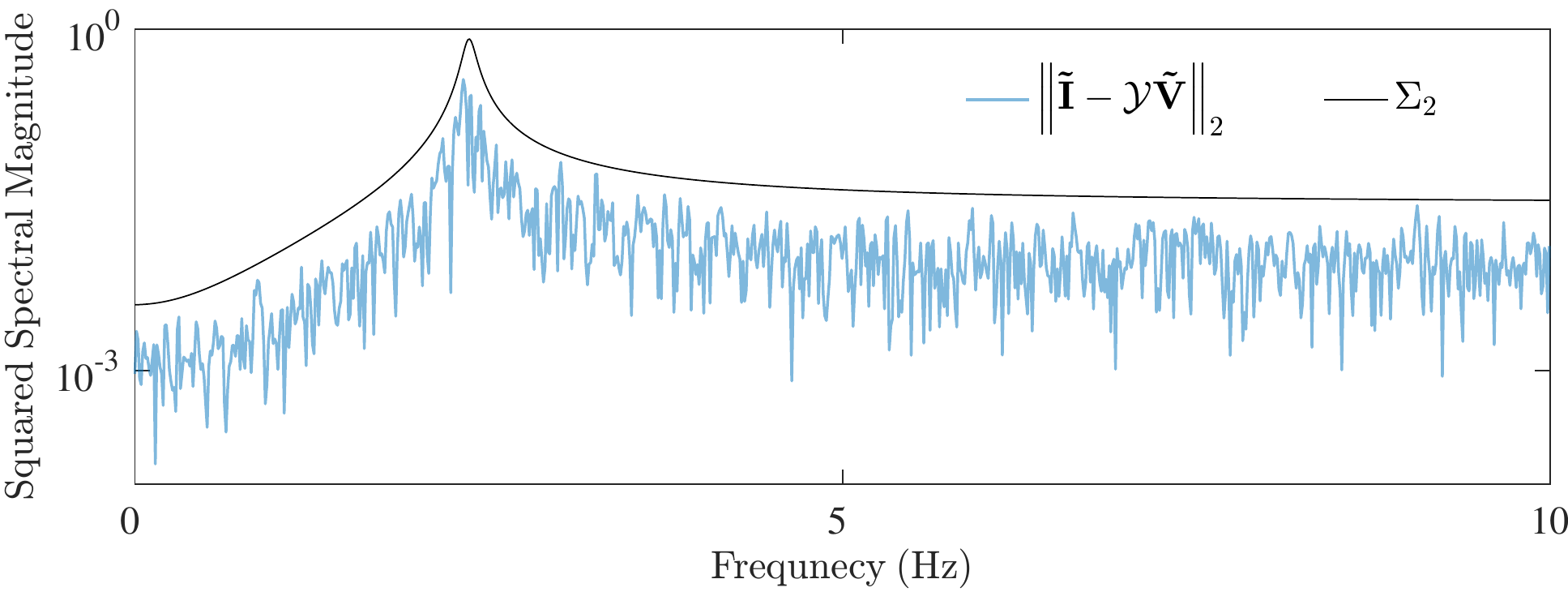}
\caption{\label{fig: LSD_Gen3} The prediction error $\Vert {\bf \tilde{I}}-\mathcal{Y}{\bf \tilde{V}}\Vert _{2}$ and the maximum measurement noise error $\Sigma_{2}$ associated with the non-source generator at bus 9 are plotted. Since there is no internal forced oscillation, prediction error is mostly caused by measurement error. Accordingly, the prediction error is bounded by the conservative maximum measurement noise error estimate $\Sigma_2$.}
\end{figure}

After simulating the system for 100s, the PMU data from each generator were collected and analyzed according to Algorithm (\ref{alg: FO}). In building the FRF of (\ref{eq: FRF}) for each generator, it was assumed generator model parameters were initially known precisely (the end of this subsection will consider parameter uncertainty). Fig. \ref{fig: Mag_Gen3} shows a sample of the simulation results associated with generator bus 9 (a non-source generator). These results show three spectral lines in each panel: (i) a measured spectrum magnitude, (ii) a predicted spectrum magnitude, and (iii) a maximum bound on the associated PMU measurement error $\Sigma_{2}$. (\ref{eq: max_l2error}) was used to compute $\Sigma_{2}$ along with the approximation given by (\ref{eq: assume_lambda}). We further assume that $\sigma^2_{\rm PMU}$ is roughly known for each PMU. Fig. \ref{fig: Mag_Gen3} shows that the measured and predicted current (phase and magnitude) spectrums begin to deviate sharply for frequencies higher than 1 Hz. This is due to the fact that the admittance matrix amplifies the mid and high frequency measurement noise, which begins the greatly dominate the voltage signal. Fig. \ref{fig: LSD_Gen3} shows that the prediction error, though, is always lower than the measurement error bound. This implies that the generator at bus 9 is not an oscillation source.

\begin{figure}
\includegraphics[scale=0.45]{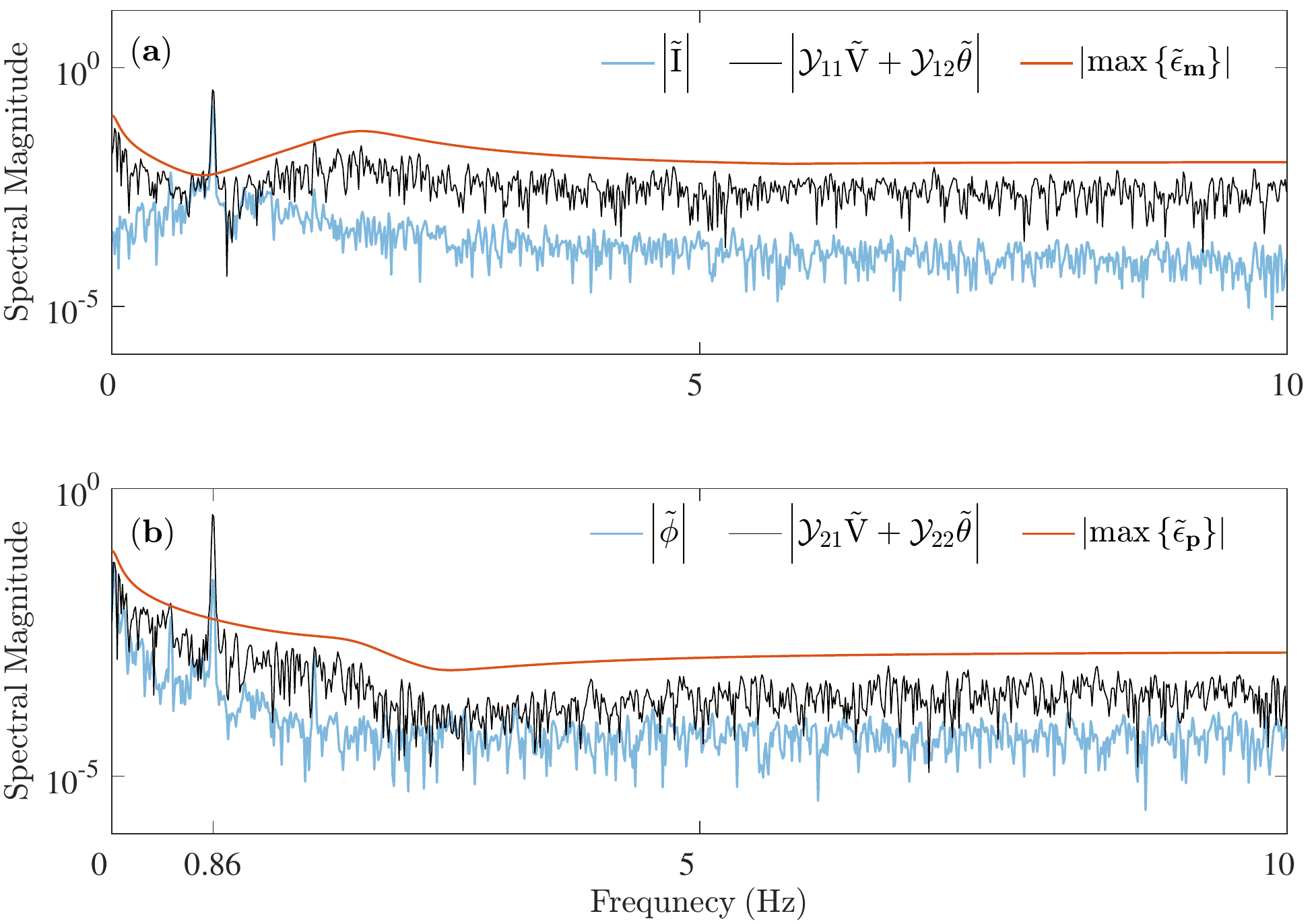}
\caption{\label{fig: Mag_Gen1} The spectral magnitude of the measured current magnitude (panel $({\bf a})$) and the measured current phase (panel $({\bf b})$) perturbations at generator bus 4 are given by the blue traces. T The associated predicted spectral magnitudes are given by the black traces. Finally, the orange traces give the estimated maximum PMU measurement noise errors.}
\end{figure}

\begin{figure}
\includegraphics[scale=0.45]{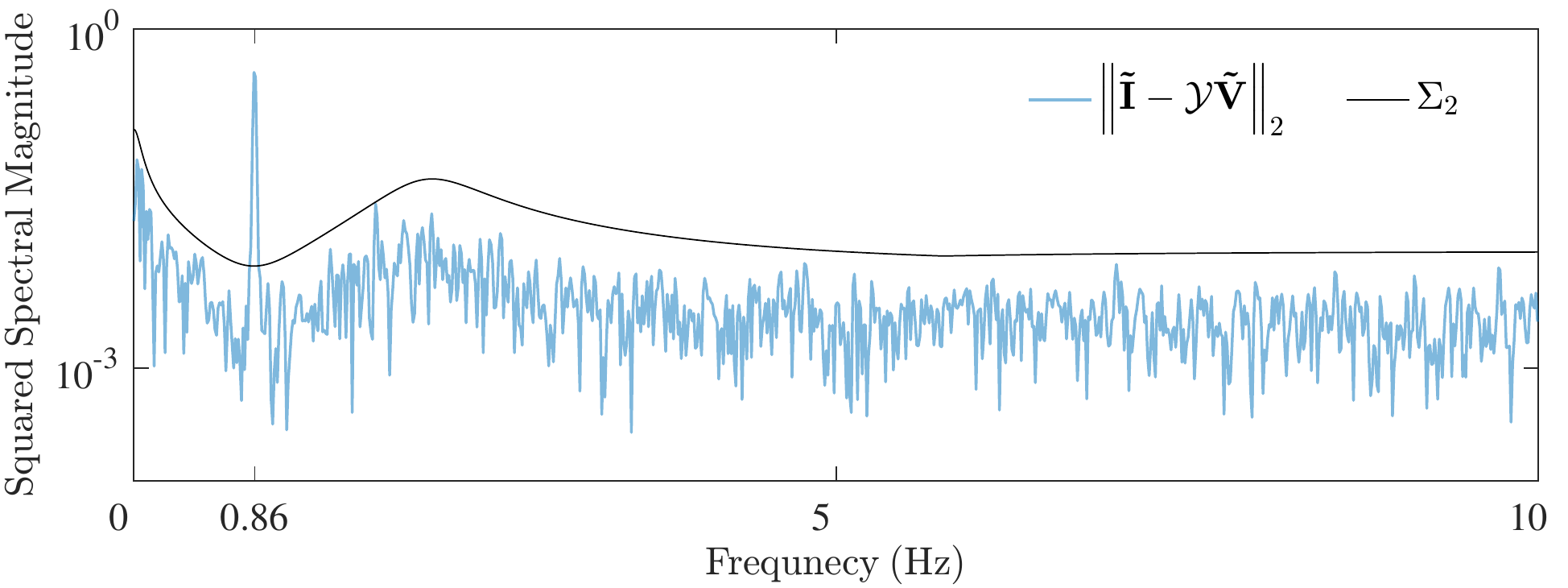}
\caption{\label{fig: LSD_Gen1} The prediction error $\Vert {\bf \tilde{I}}-\mathcal{Y}{\bf \tilde{V}}\Vert _{2}$ and the maximum measurement noise error $\Sigma_{2}$ associated with the source generator at bus 4 are plotted. Since there is an internal forced oscillation at $f_d=0.86$ Hz, the prediction error greatly exceeds the measurement noise error bound at this frequency.}
\end{figure}

\begin{figure}
\includegraphics[scale=0.45]{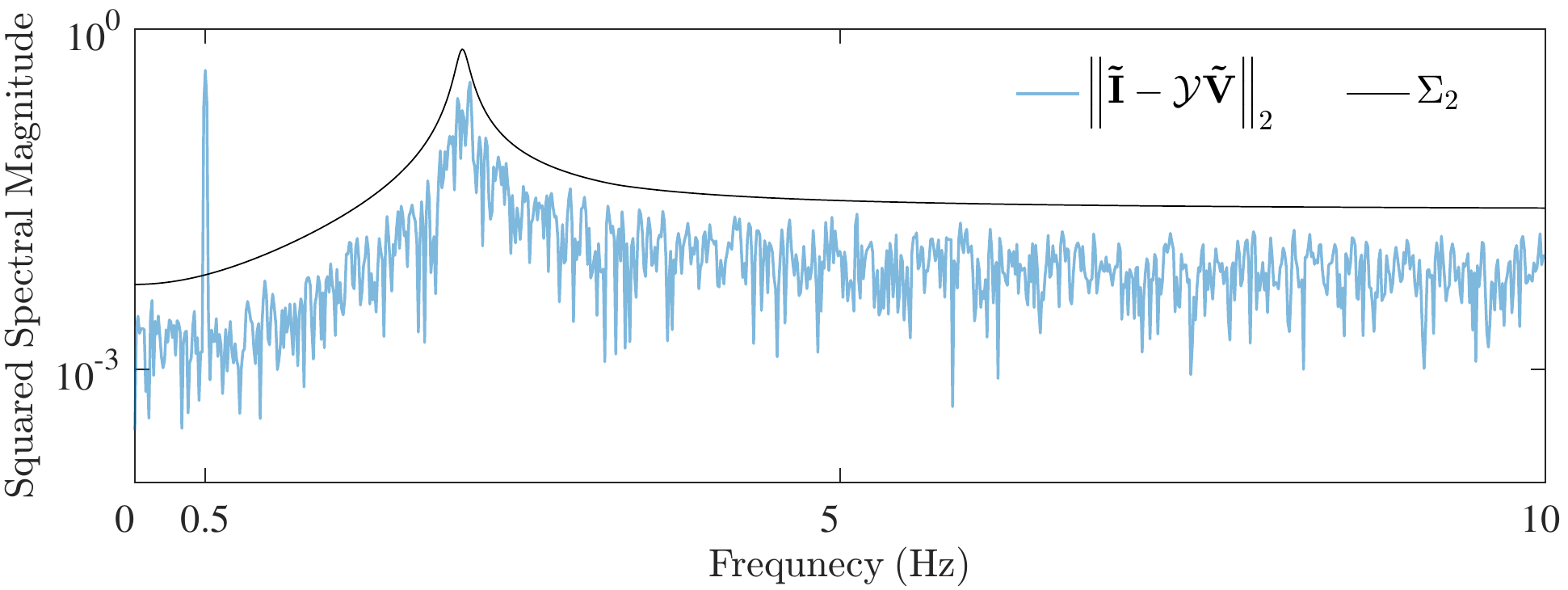}
\caption{\label{fig: LSD_Gen5} The prediction error $\Vert {\bf \tilde{I}}-\mathcal{Y}{\bf \tilde{V}}\Vert _{2}$ and the maximum measurement noise error $\Sigma_{2}$ associated with the source generator at bus 13 are plotted. Since there is an internal forced oscillation at $f_d=0.5$ Hz, the prediction error greatly exceeds the measurement noise error bound at this frequency.}
\end{figure}

\begin{figure}
\includegraphics[scale=0.45]{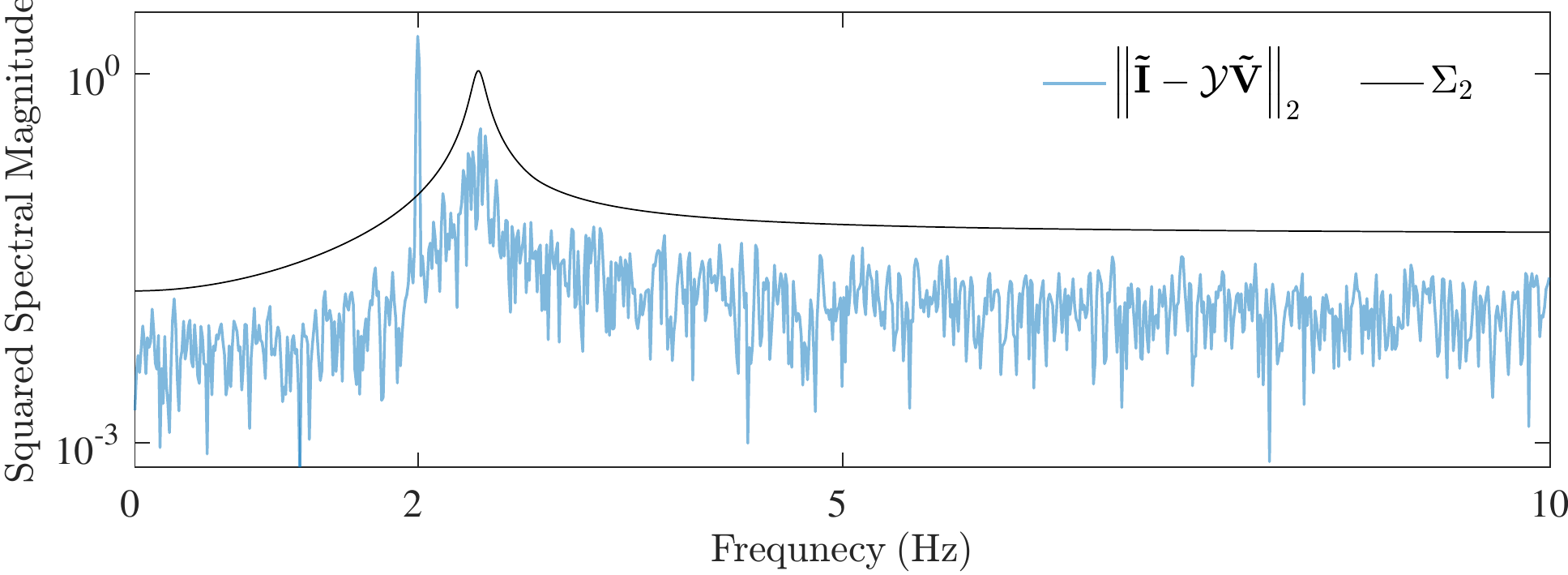}
\caption{\label{fig: LSD_Gen15} The prediction error $\Vert {\bf \tilde{I}}-\mathcal{Y}{\bf \tilde{V}}\Vert _{2}$ and the maximum measurement noise error $\Sigma_{2}$ associated with the source generator at bus 65 are plotted. Since there is an internal forced oscillation at $f_d=2.0$ Hz, the prediction error greatly exceeds the measurement noise error bound at this frequency.}
\end{figure}

The results of Figs. \ref{fig: Mag_Gen3} and \ref{fig: LSD_Gen3}, which are associated with a non-source bus, can be contrasted to Figs. \ref{fig: Mag_Gen1} and \ref{fig: LSD_Gen1}, which are associated with source bus 4. At this generator, the AVR reference is oscillated at $f_d=0.86$ Hz. This causes large observable differences in the measured and predicted magnitude spectrums. In Figs. \ref{fig: LSD_Gen5} and \ref{fig: LSD_Gen15}, the prediction error and measurement noise error bounds are also contrasted at generators 13 and 65 (both source generators). As can be seen, there is significant spectral error at the forcing frequencies which the measurement noise cannot account for. This implies that both of these generators are sources of forced oscillations.

\begin{figure}
\includegraphics[scale=0.45]{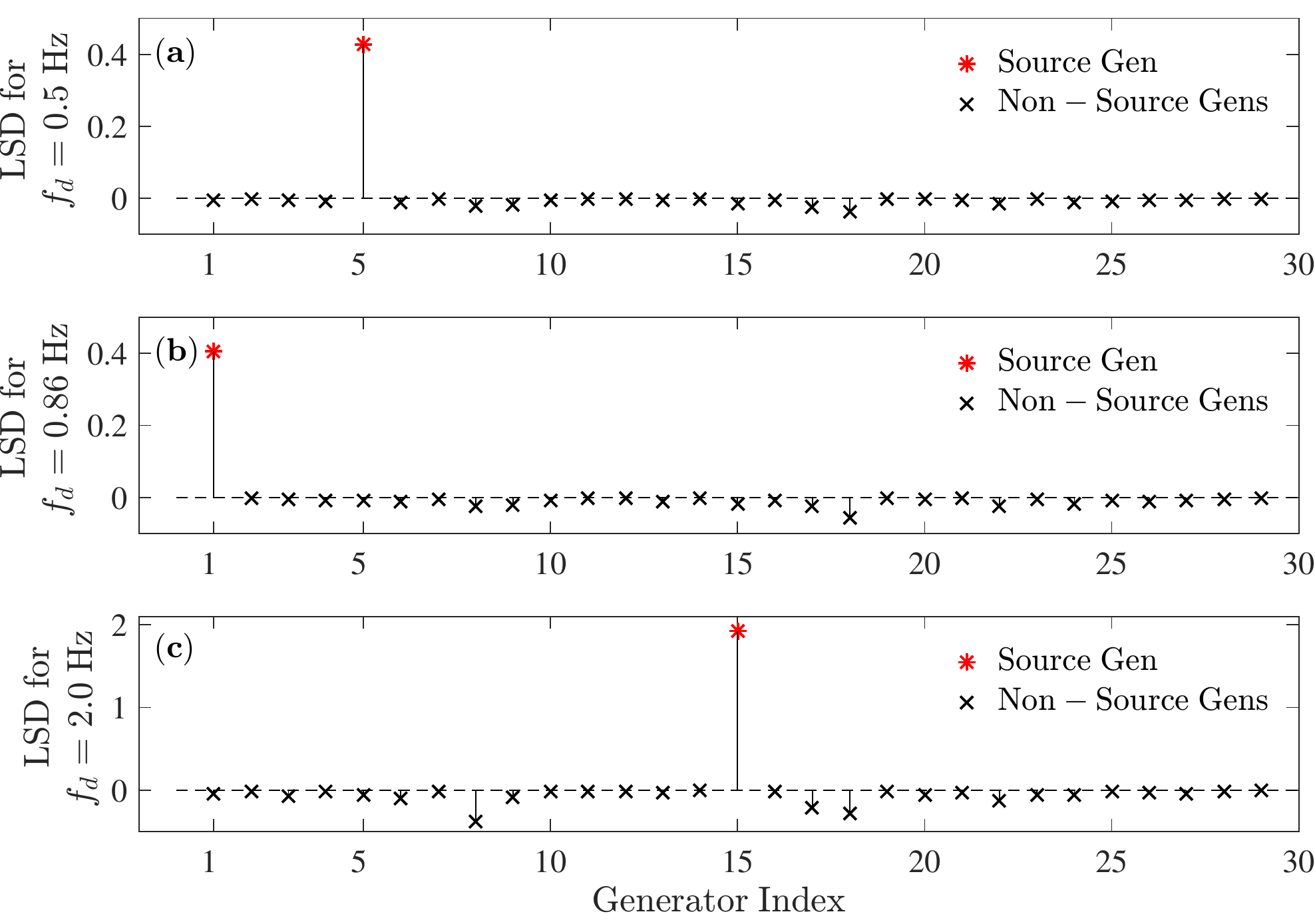}
\caption{\label{fig: LSD_Indices} The LSD is computed at each generator for $f_d=0.5$ Hz (panel $({\bf a})$), $f_d=0.86$ Hz (panel $({\bf b})$), and $f_d=2.0$ Hz (panel $({\bf c})$). At each frequency, the correct generator is located, despite strong PMU measurement noise. Generator index 1 corresponds to the generator at bus 4, generator index 5 corresponds to the generator at bus 13, and generator index 15 corresponds to the generator at bus 65.}
\end{figure}

After analyzing the generator spectrums, the LSD can be quantified at each forcing frequency across all 29 system generators. These results are given in Fig \ref{fig: LSD_Indices}. In plotting the LSD indices for each generator at each forcing frequency, the largest spectral deviations are easily found at the correct source generators. We do not formally define a threshold parameter $\iota$, which is required in the final steps of Algorithm (\ref{alg: FO}), since it would have to be found empirically, by a system operator, via PMU data collected over time. We currently do not have access to such data.

\begin{figure}
\includegraphics[scale=0.45]{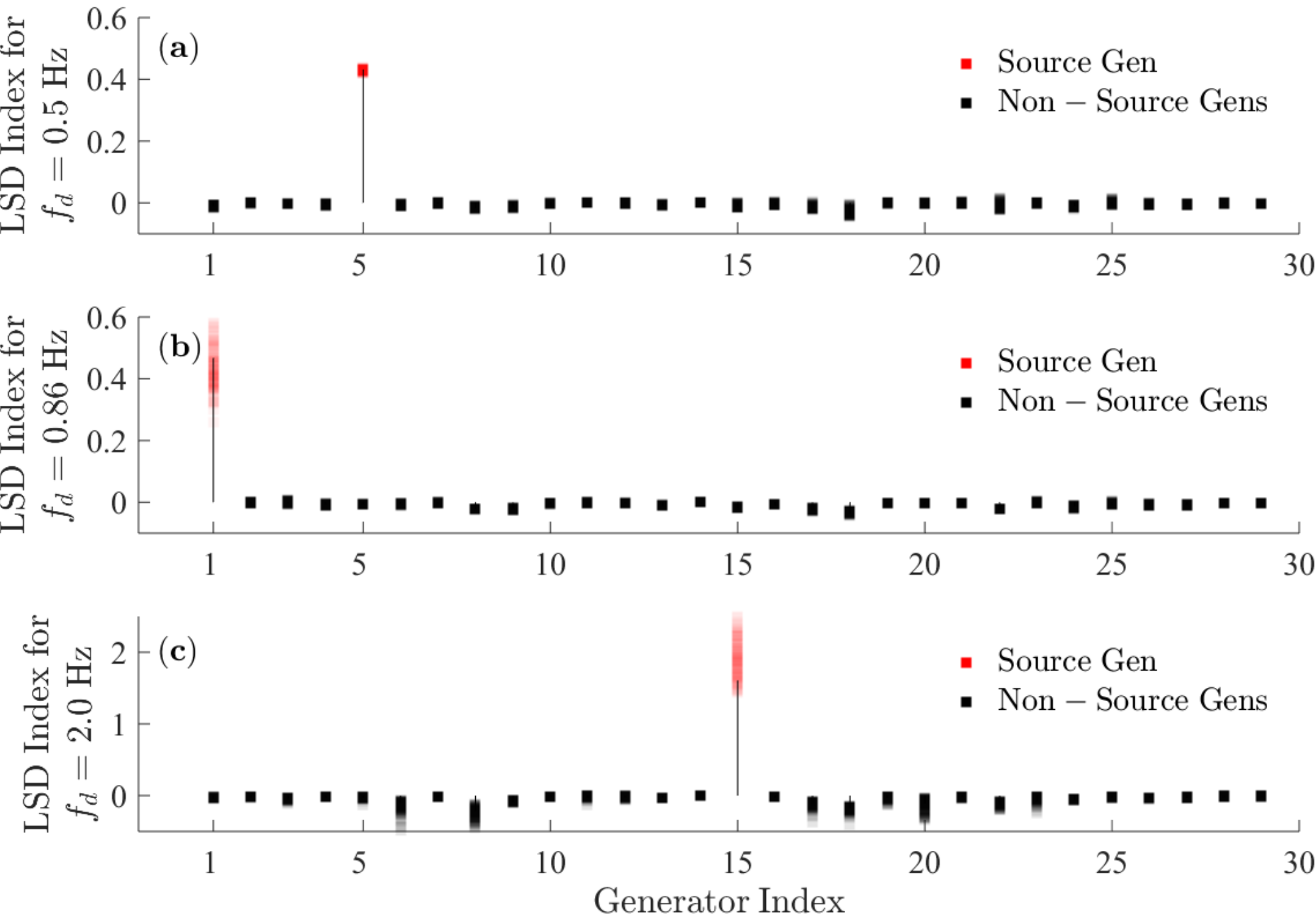}
\caption{\label{fig: LSD_Indices_Params} The LSD is computed at each generator for $f_d=0.5$ Hz (panel $({\bf a})$), $f_d=0.86$ Hz (panel $({\bf b})$), and $f_d=2.0$ Hz (panel $({\bf c})$) over 100 trials to consider the impact of generator parameter uncertainty.}
\end{figure}

Although system generators may be modeled reasonably accurately, the generator model parameters may be known to a lesser degree of accuracy. To consider the effects of generator parameter uncertainty, the LSD is re-quantified for each generator, but in building the FRF of (\ref{eq: FRF}), generator parameter uncertainty is introduced over 100 trials. Parameter uncertainty includes all damping, reactance, time constant, and AVR variables. Inertia uncertainty is not considered since this is a static and typically very well defined parameter. All parameters are altered by a percentage chosen from a normal distribution characterized by $\mu=0$ and $\sigma=10$\%, meaning uncertainty can range up to $\pm30\%$. This was the largest standard deviation for which parameters uncertainty was tolerable. The results, given by Fig. \ref{fig: LSD_Indices_Params}, show that the LSD metric is fairly robust to model parameter uncertainty, although future work will refine this method for enhanced accuracy. In general, this parameter uncertainty analysis indicates that a reasonably accurate generator model is necessary to employ these frequency response methods at any particular generator.

\subsection{WECC-179 Bus System (Natural Oscillation)}\label{sbsec: 179_NO}
\begin{figure}
\includegraphics[scale=0.45]{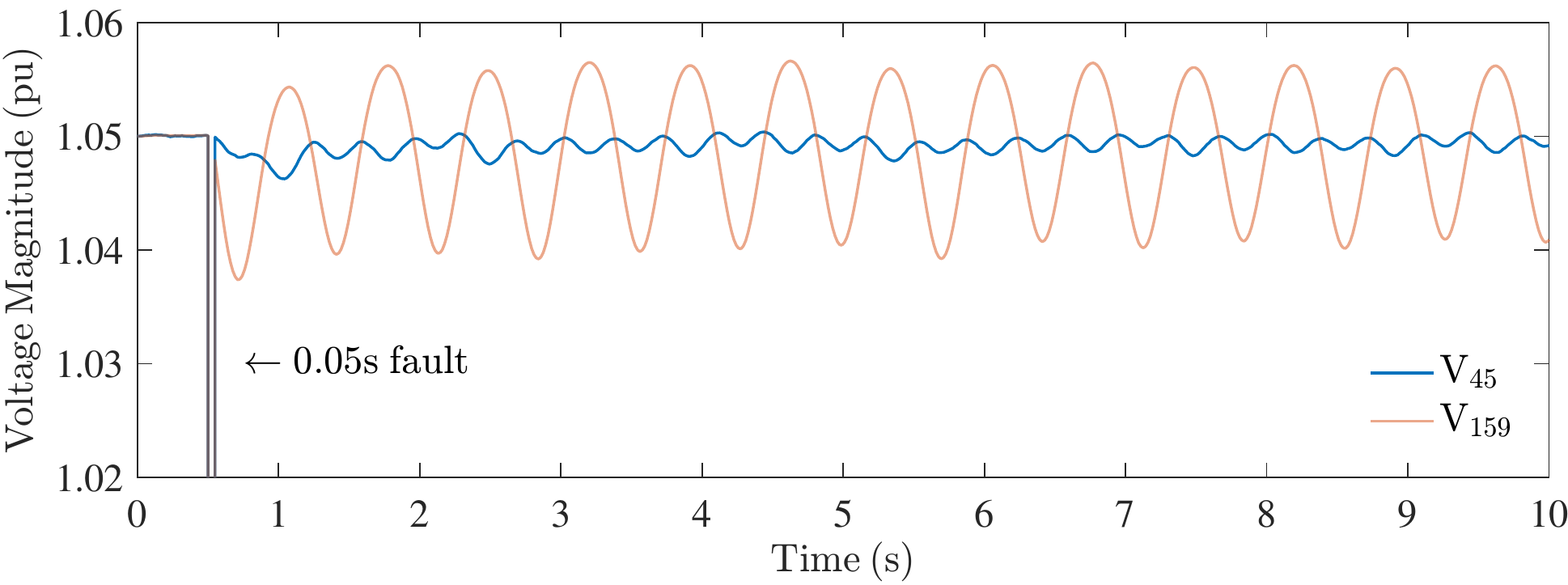}
\caption{\label{fig: ND_Vts} The voltage magnitude at buses 45 and 159 are plotted before, while, and briefly after the system experiences a fault. Measurement noise is not shown.}
\end{figure}

As a third test, the admittance matrix source location technique was applied in the presence of a natural oscillation (no forced oscillation sources). We used test case ``ND1" from~\cite{Maslennikov:2016}, where a natural oscillation is excited in the WECC 179-bus system. In ``ND1", all generators are modeled as $2^{\rm nd}$ order, and most are assigned a damping parameter of $D=4$. The generators at buses 45 and 159, though, are assigned $D=-1.5$ and $D=1$, respectively, such that there exists a poorly damped mode with damping ratio $\zeta=0.01$. To excite the system's underdamped mode, a fault is applied at bus 159 for 0.05s. This system was simulated for 100 seconds with the same load dynamics and PMU measurement noise assumptions as were used in simulating test case ``F1". The bus voltage magnitude from generator buses 45 and 159 (oscillations are strongest at these generators) are given before, during, and after the fault by Fig. \ref{fig: ND_Vts}. As can be inferred from this plot, the excited underdamped natural mode of this system has frequency $f_n=1.41$ Hz.

\begin{figure}
\includegraphics[scale=0.45]{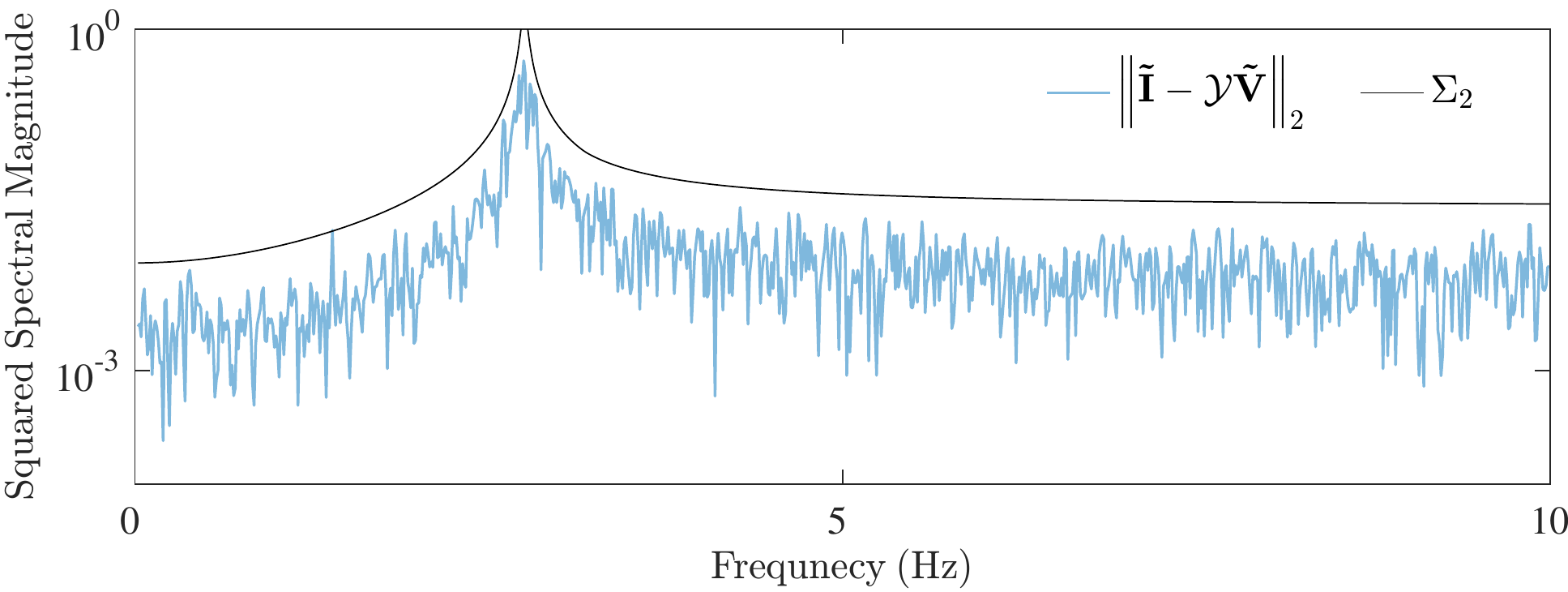}
\caption{\label{fig: LSD_Gen13_ND} The prediction error $\Vert {\bf \tilde{I}}-\mathcal{Y}{\bf \tilde{V}}\Vert _{2}$ and the noise error bound $\Sigma_{2}$ associated with generator 45 are plotted for test case ND1. The prediction error slightly exceeds the noise error bound at $f=1.41$ Hz.}
\end{figure}

\begin{figure}
\includegraphics[scale=0.45]{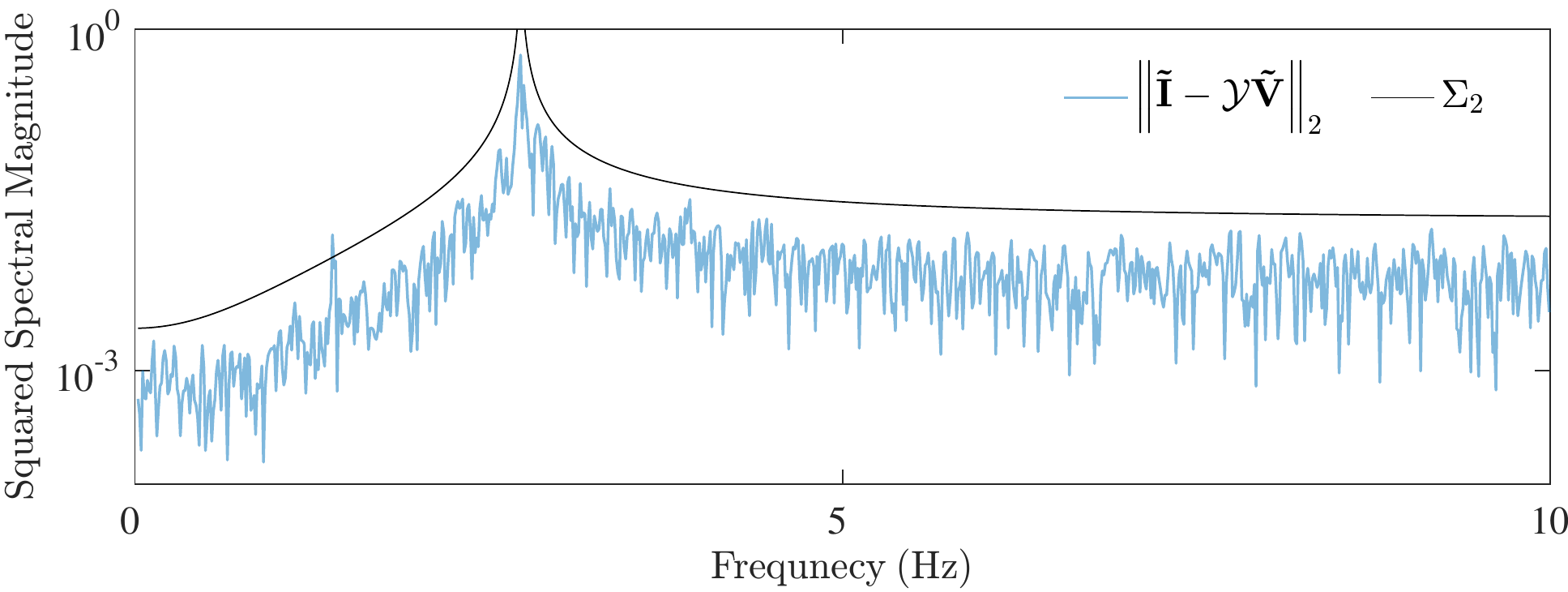}
\caption{\label{fig: LSD_Gen28_ND} The prediction error $\Vert {\bf \tilde{I}}-\mathcal{Y}{\bf \tilde{V}}\Vert _{2}$ and the noise error bound $\Sigma_{2}$ associated with generator 159 are plotted for test case ND1. The prediction error slightly exceeds the noise error bound at $f=1.41$ Hz.}
\end{figure}

Since the persisting oscillations are caused by the excitation of a poorly damped mode, we say the system is experiencing a natural oscillation rather than a forced oscillation. Therefore, the source location technique should indicate that no generator contains an internal forcing function. To test this theory, the prediction error aggregate $\Vert {\bf \tilde{I}}-\mathcal{Y}{\bf \tilde{V}}\Vert _{2}$ and the noise error bound $\Sigma_{2}$ were calculated via Algorithm (\ref{alg: FO}) and plotted for generator buses 45 and 159 (see Figs. \ref{fig: LSD_Gen13_ND} and \ref{fig: LSD_Gen28_ND}, respectively). In each of these cases, the prediction error slightly exceeds the noise error at $f_n=1.41$ Hz. This deviation is very small, though, relative to the strength of the oscillation, and is likely due to slight nonlinearity of the generator responses (generator current angular perturbations are very large).

\begin{figure}
\includegraphics[scale=0.44]{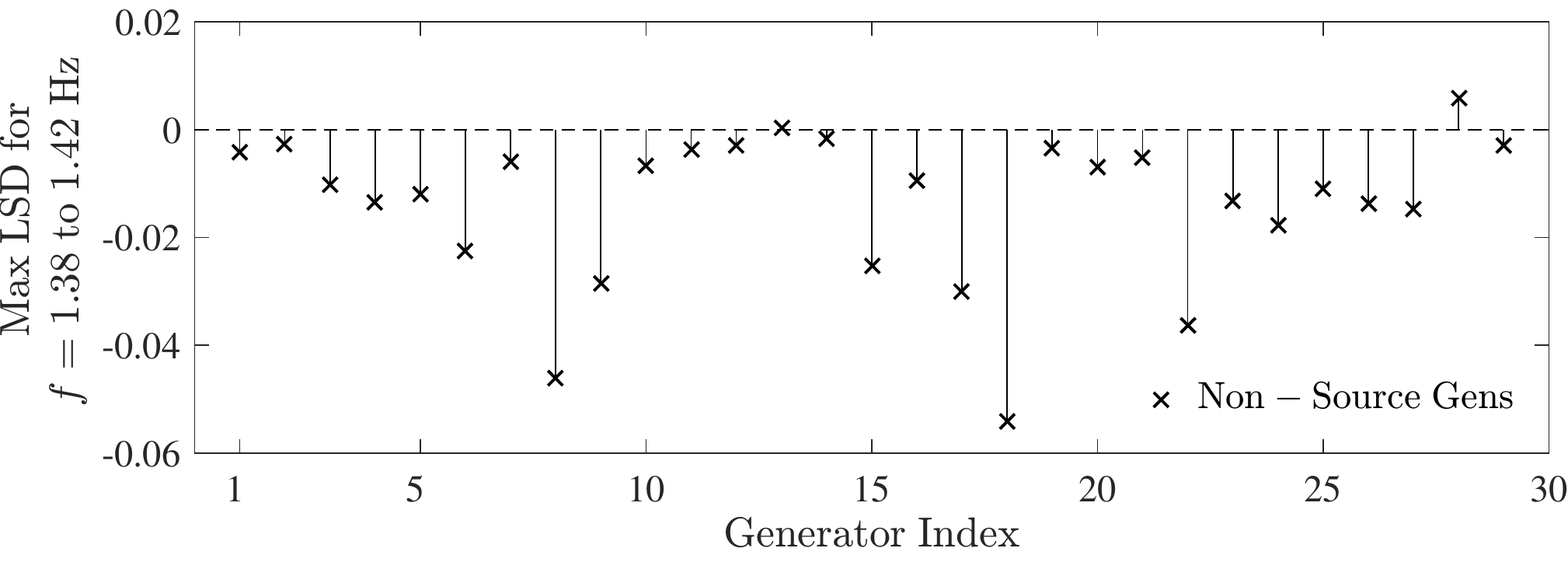}
\caption{\label{fig: LSD_Indices_ND} The maximum LSD, from $f=1.38$ to $f=1.42$, is plotted for each generator. The maximum LSDs at generators 13 (bus 45) and 28 (bus 159) are slightly positive, but are still sufficiently small.}
\end{figure}

To further analyze the system, the LSDs were calculated at each generator (we again assumed PMU data were available). Since the LSD is a function of frequency, and there is no forcing frequency, we computed the LSDs at all generators in the range of $f=1.38$ to $f=1.42$ Hz. We then plot the maximum LSD in this frequency band for each generator. This result is shown in Fig. \ref{fig: LSD_Indices_ND}. In this plot, the maximum LSDs at generators 13 (bus 45) and 28 (bus 159) are seen to cross the zero threshold. Given the strength of the oscillation, as seen in Fig. \ref{fig: ND_Vts}, and the very small deviation between the prediction and the measurement, none of the sampled generators could be forced oscillation source candidates. More formally, all calculated LSD values are smaller than any realistically chosen $\iota$ parameter which would represent the threshold for determining if a generator is the source of a forced oscillation. We may thus conclude that either the system is being forcibly oscillated by some non-generator piece of equipment or load, or that a natural oscillation is driving the system's periodic dynamics.

\subsection{WECC-179 Bus System (Forced + Natural Oscillation)}\label{sbsec: 179_FNO}
As a fourth test case, we used the natural oscillation test case ``ND2'' and we added the forced oscillation described in test case ``F63"  (both are described in~\cite{Maslennikov:2016}). Specifically, we set the damping parameters of the generators at buses 35 ($D_{35}=0.5$) and 65 ($D_{65}=-1$) such that there exists a poorly damped mode ($\zeta = 0.02\%$) at 0.37 Hz. Additionally, we forcibly oscillated generator 79's AVR reference voltage with a additive \textit{square wave} of frequency 0.40 Hz. In this particular situation, the presence of a negative damping at generator 65 can cause the generator to be viewed as a source of the so called ``transient energy" in the DEF method. Accordingly, the DEF method will locate this generator as the source of the negative damping. Our FRF method, though, may be used in a complimentary fashion to find the forced oscillation source.

\begin{figure}
\includegraphics[scale=0.45]{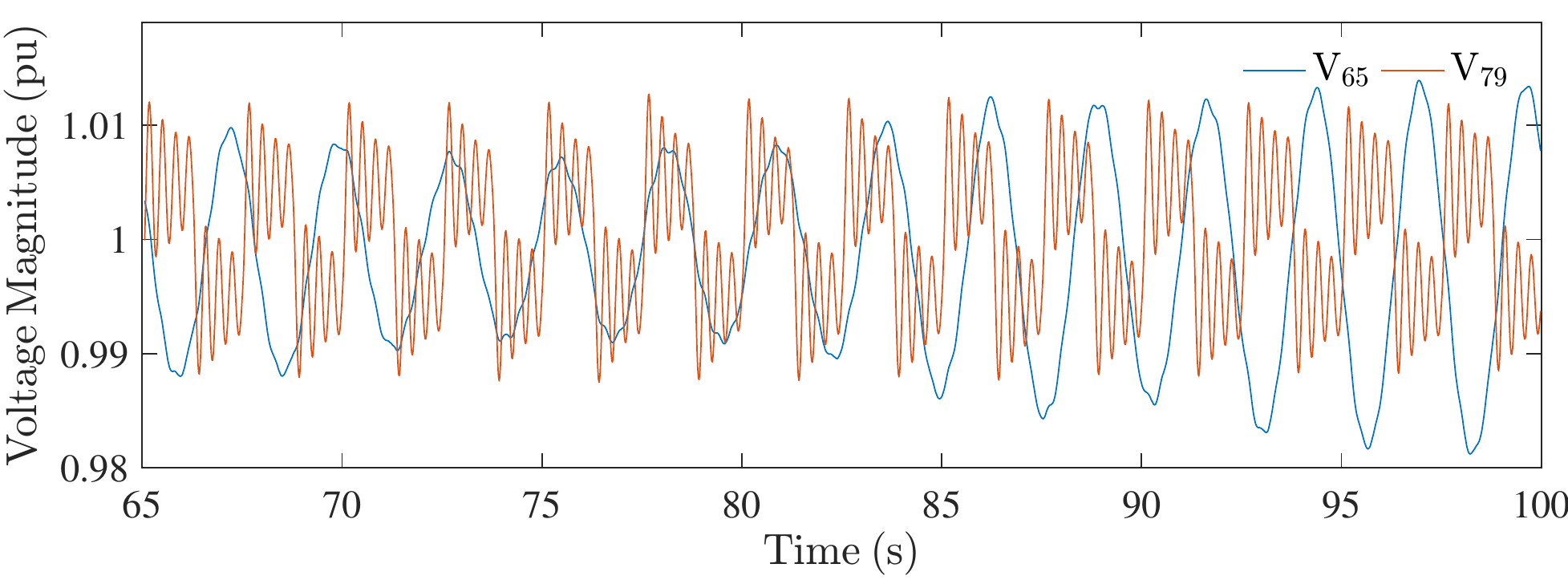}
\caption{\label{fig: Vm_NFO} The voltage magnitude at buses 65 and 79 are plotted over 35 seconds. The natural mode frequency of 0.37 Hz, the forcing frequency of 0.40 Hz, and the resulting beat frequency can all be seen clearly. Measurement noise is not shown.}
\end{figure}

\begin{figure}
\includegraphics[scale=0.45]{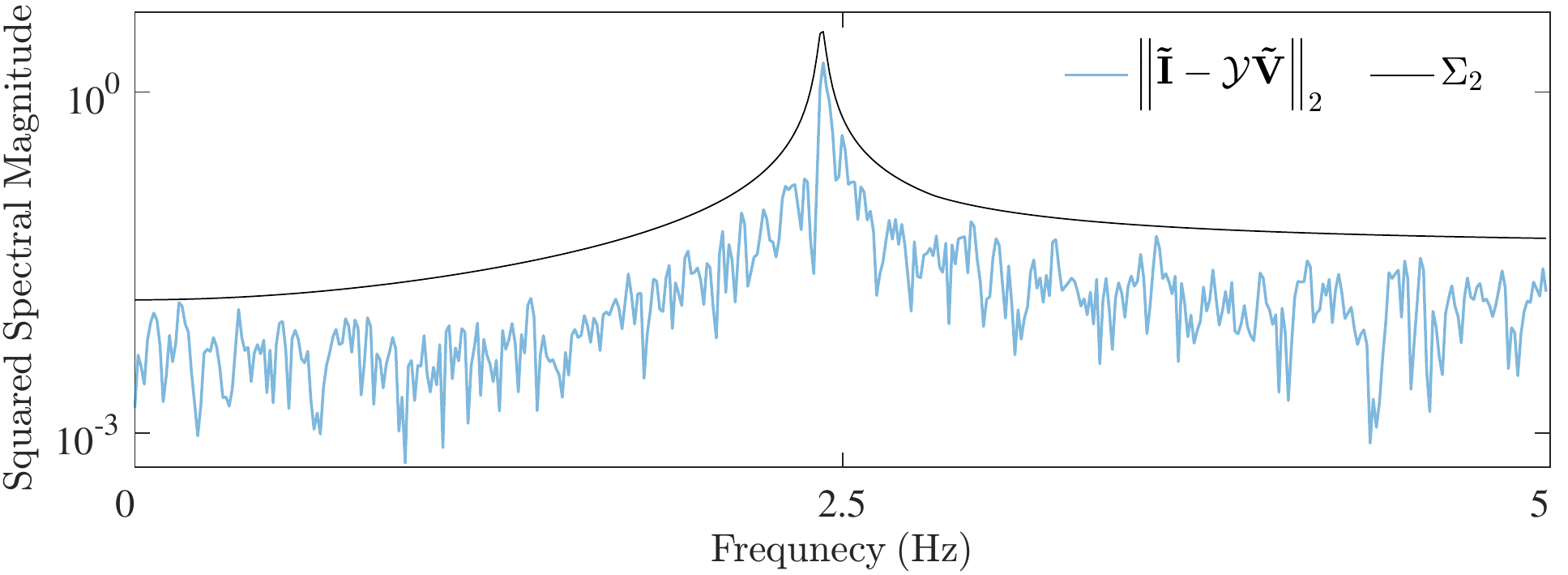}
\caption{\label{fig: LSD_Gen15_NDF} The prediction error $\Vert {\bf \tilde{I}}-\mathcal{Y}{\bf \tilde{V}}\Vert _{2}$ and the noise error bound $\Sigma_{2}$ associated with generator 65 are plotted for the test case where an underdamped natural mode is excited by a forced oscillation. Despite a strong oscillatory response from generator 65 at 0.37 Hz, the prediction error is entirely contained by the measurement noise error bound for all frequencies.}
\end{figure}

\begin{figure}
\includegraphics[scale=0.45]{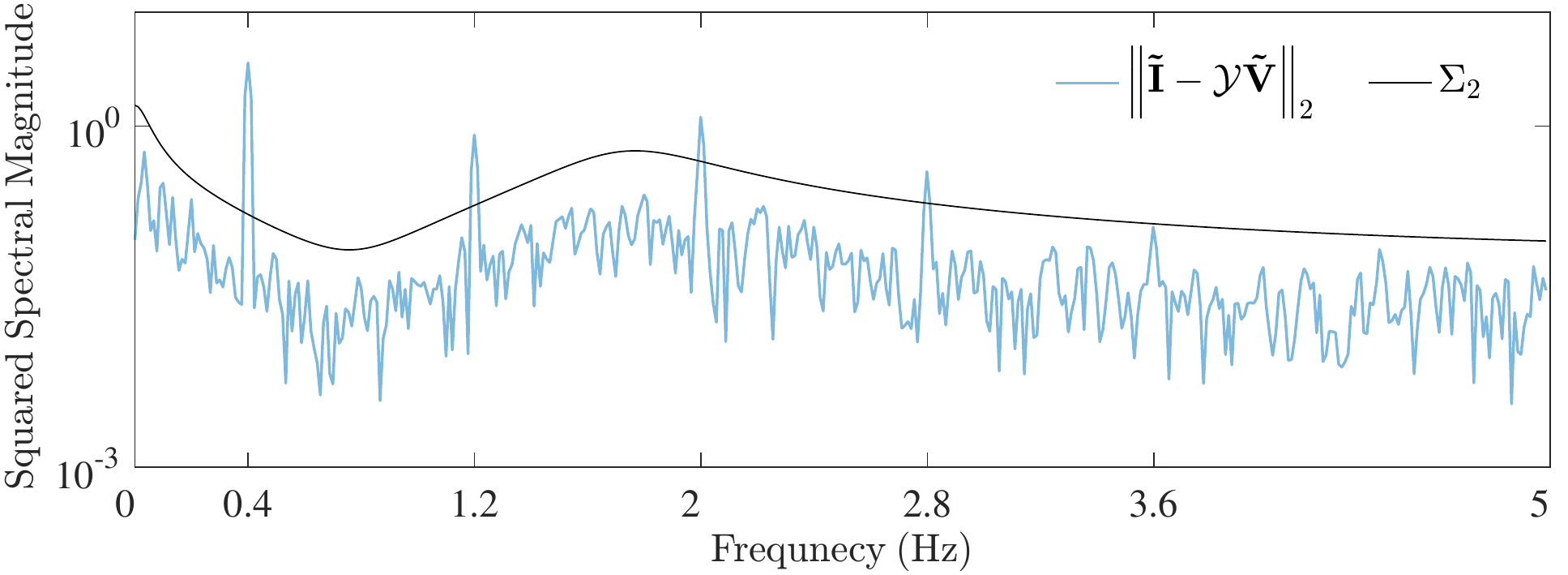}
\caption{\label{fig: LSD_Gen18_NDF} The prediction error $\Vert {\bf \tilde{I}}-\mathcal{Y}{\bf \tilde{V}}\Vert _{2}$ and the noise error bound $\Sigma_{2}$ associated with generator 79, the source bus, are plotted for the test case where an underdamped natural mode is excited by a forced oscillation. The prediction error violates the measurement noise error at $f=0.4, \; 1.2, \; 2.0$ and $2.8$ Hz.}
\end{figure}

\begin{figure}
\includegraphics[scale=0.45]{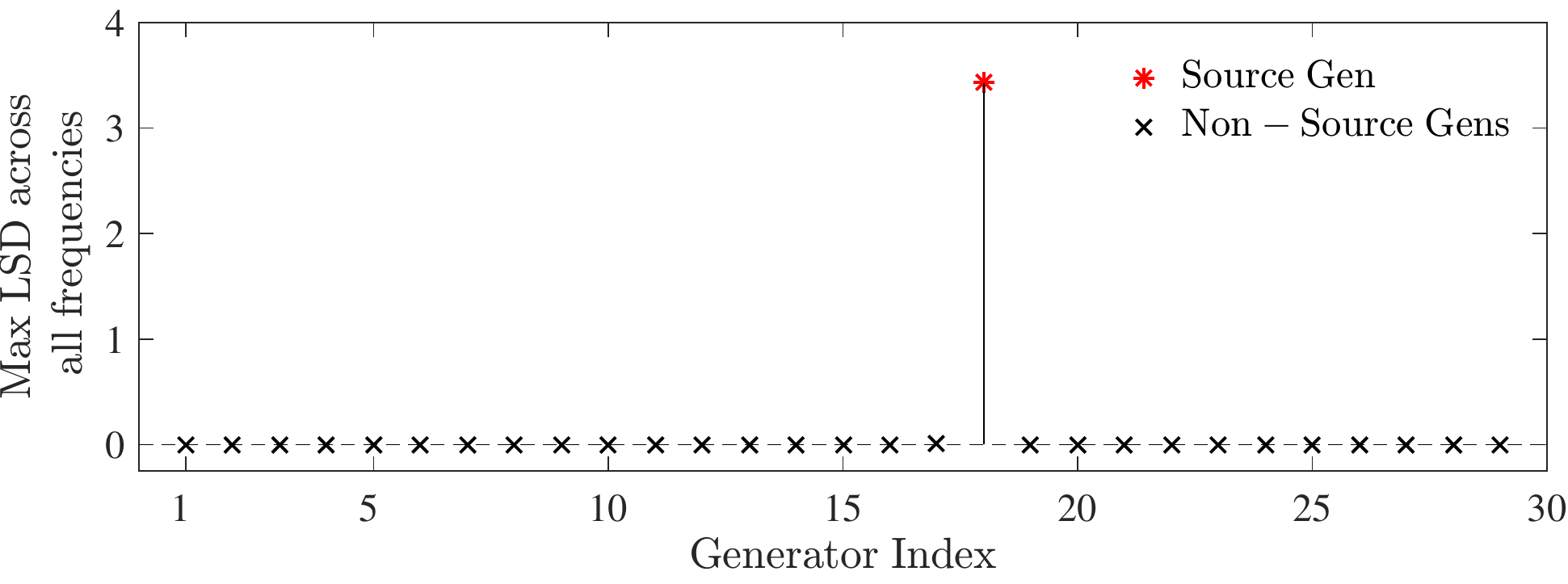}
\caption{\label{fig: LSD_Indices_NFD} The LSD is computed at each generator across the full range of measured frequencies for the natural + forced oscillation test case. Only the largest LSD for each generator is plotted here, though. Generator index 18, which corresponds to the generator at bus 79, is correctly identified as the source generator.}
\end{figure}

The voltage magnitudes at buses 65 and 79 are shown in Fig. \ref{fig: Vm_NFO} over 35 seconds. Generator 79's response to the additive square wave on the AVR reference is evident. In this test, the forced oscillation frequency is only slightly larger than the natural frequency of the poorly damped mode. This elicits a strong response from the generator at bus 65. Accordingly, we compare the prediction error and measurement noise bound at both generators across the full spectrum of frequencies. In Fig. \ref{fig: LSD_Gen15_NDF}, the the prediction error is seen to be totally contained by the measurement noise error bound at generator 65. This is true for all other generators (aside from generator 79) in the system as well. The resulting negative LSDs at all of these generators, across all frequencies, along with the massively positive LSD at generator 79, indicates there is only one forced oscillation source. This is shown by Fig. \ref{fig: LSD_Indices_NFD}. Further evidence that generator 79 is the source of the oscillation can be seen by the Fig. \ref{fig: LSD_Gen18_NDF}. There are a series of prediction error spikes which violate the measurement noise error bound. The statistical signatures of these spikes further indicate that the forcing function is a square wave. To understand why, equation (\ref{eq: fs_SW}) gives the Fourier series of a pure square wave $g_s(t)$ with fundamental frequency $f$. This series contains frequencies $f$, $3f$, $5f$, and so on, just as spectral deviations in Fig. \ref{fig: LSD_Gen18_NDF} occur at $f=0.4, \; 1.2, \; 2.0$ and $2.8$ Hz:
\begin{align}\label{eq: fs_SW}
g_{s}(t)=\frac{4}{\pi}\sum_{n=1,3,5...}^{\infty}\frac{\sin\left(2\pi nft\right)}{n}.
\end{align}

\subsection{3-Bus System with Constant Impedance}\label{sbsec: Const_Imp}
\begin{figure}
\begin{centering}
\includegraphics[scale=1]{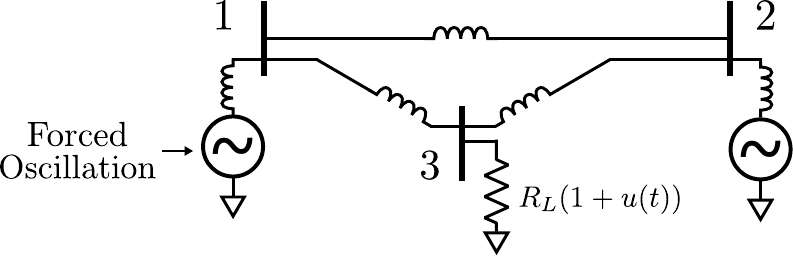}
\par\end{centering}
\caption{\label{fig: 3_Bus_DEF} 3 bus system with two $2^{\rm nd}$ order generators and a resistive load. Resistive Ornstein-Uhlenbeck noise is added to mimic system fluctuations.}
\end{figure}

As indicated in~\cite{Chen:2017}, network resistances embedded in system transfer conductances (shunt and series) and constant impedance loads may act as the source of transient energy from the viewpoint of the DEF method. The simplest system known to exhibit this phenomena~\cite{Chen:2017} can be modeled as a two generator system with some constant impedance load (or shunt), as given by Fig. \ref{fig: 3_Bus_DEF}. In this system, we apply light Ornstein-Uhlenbeck noise of (\ref{eq: u_dot}) to the resistive load in order to mimic system fluctuations, and we apply a forced oscillation of $\Omega_d=2\;\frac{\rm rad}{\rm sec}$ to the torque on generator 1.

\begin{figure}
\includegraphics[scale=0.44]{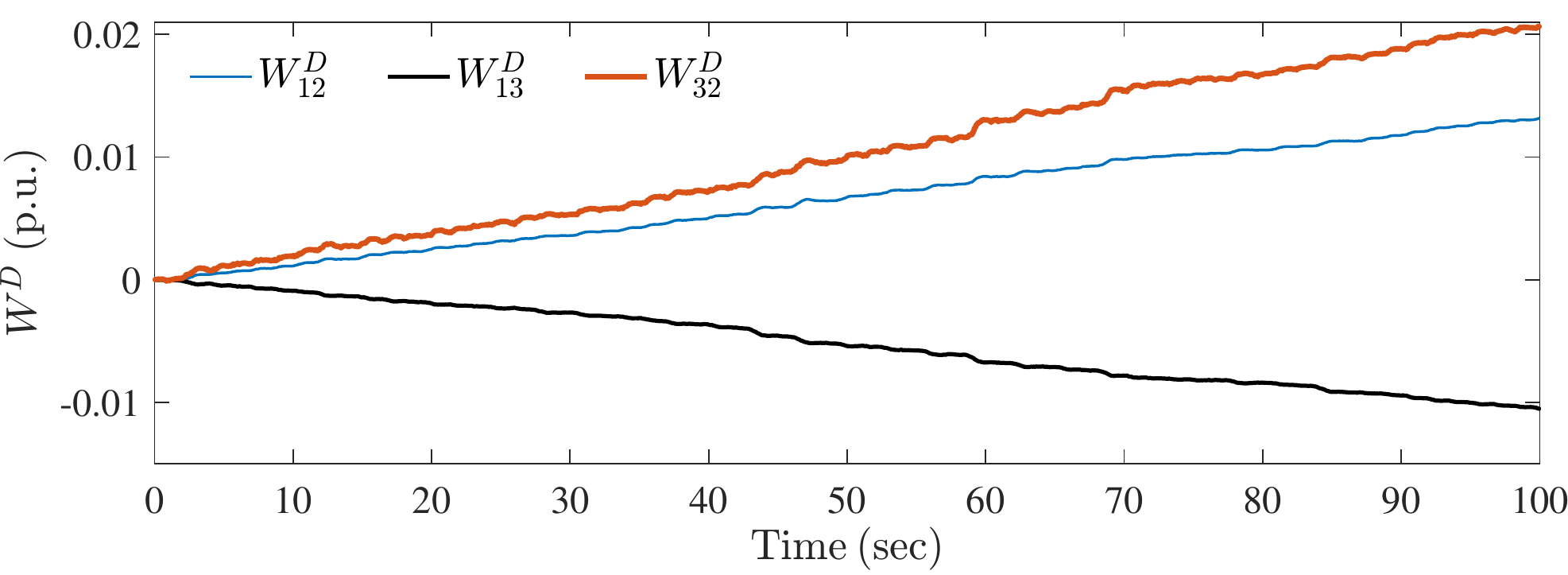}
\caption{\label{fig: WD_DEF} The DEF is computed for lines \{12\}, \{13\}, and \{32\}.}
\end{figure}

\begin{figure}
\includegraphics[scale=0.45]{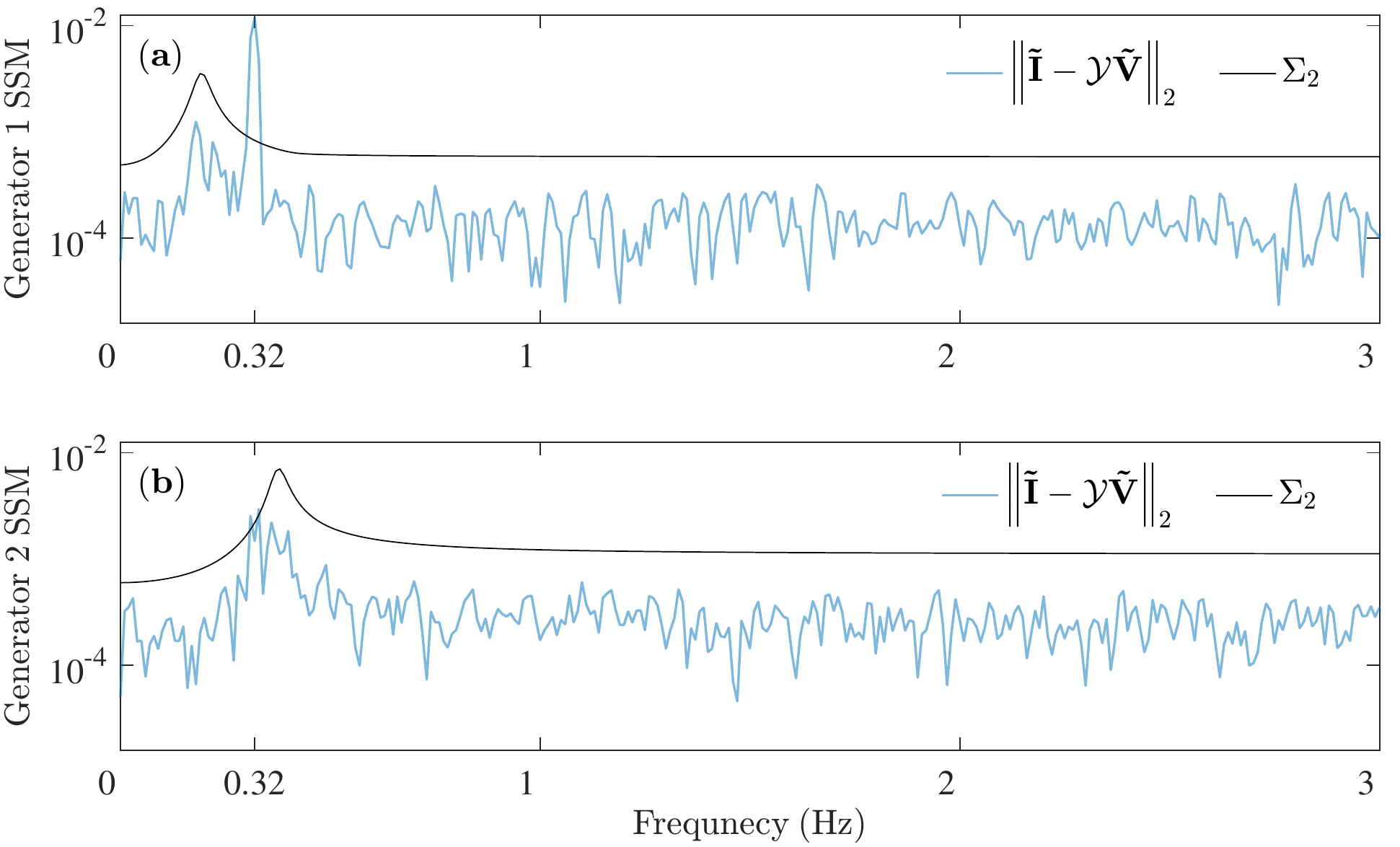}
\caption{\label{fig: Spec_Dev_Def} The prediction error $\Vert {\bf \tilde{I}}-\mathcal{Y}{\bf \tilde{V}}\Vert _{2}$ (given as the Squared Spectral Magnitude (SSM)) and the noise error bound $\Sigma_2$ associated with generator 1 (panel $({\bf a})$) and generator 2 (panel $({\bf b})$) are given.}
\end{figure}

After simulating this system and adding white PMU measurement noise with $\sigma=0.1$ ($\%\; {\rm pu}$), the flow of dissipating energy was computed according to~\cite[eq. (3)]{Maslennikov:2017}. The results are given by Fig. \ref{fig: WD_DEF}. According to the notation introduced in~\cite[eq. (5)]{Maslennikov:2017}, we found that $DE^*_{12}=0.61$, $DE^*_{13}=-0.48$, and $DE^*_{32}=0.94$. These results indicate that energy is flowing from the resistive load at bus 3 to the two generator buses. Energy is also flowing from the generator 1 (the source bus) to generator 2 (the system sink).
These results do not accurately locate the source of the oscillation due to the resistive load. The reasons why are explained in~\cite{Chen:2017} and shall not be investigated here. We then applied the FRF method to both generators. In building the FRF of $\mathcal Y$, reactance and damping parameters were perturbed by a percentage pulled from a normal distribution with standard deviation $\sigma=0.05\%$. The FO is clearly located at generator 1 due to the significantly positive LSD at 0.32Hz in panel $({\bf a})$ of Fig. \ref{fig: Spec_Dev_Def}. Conversely, the LSD at 0.32 Hz on generator 2 is effectively 0. Since the FRF method presented in this paper is invariant to network dynamics, it is not constrained by load modeling assumptions.


\section{Conclusion and Future Work}\label{Conclusion}
We have developed a method for using generator terminal PMU data to determine the source of a forced oscillation. This is accomplished by building the Frequency Response Function (FRF) for a given generator and comparing its measured and predicted current spectrums. The FRF can be derived from any arbitrary generator model without simplification, so it is thus unconstrained by model order reduction necessities. Unique measurement noise considerations are taken into account to determine if measurement and prediction deviations are due to noise or an internal forcing function. Similar to the hybrid methods of~\cite{Wu:2012} and~\cite{Ma:2010}, our method assumes prior knowledge of generator models. Unlike the hybrid models though, our method is simulation free and algebraically simple to implement. Also, PMU noise considerations are more straightforward to handle and results may be interpreted more intuitively. Through the examples provided in Section \ref{Test_Results}, we have shown that the method is robust to model parameter uncertainty, meaning that very accurate system parameter knowledge is not a binding requirement. In subsequent work, we hope to leverage this technique and the properties of the derived admittance matrices to further characterize how oscillations propagate through the transmission grid. This will lend additional understanding into the mechanisms behind the successful Dissipating Energy Flow method of~\cite{Maslennikov:2017} and provide a framework for improvement investigations.
\bibliographystyle{ieeetr}
\bibliography{FO_Bib}
\begin{IEEEbiography}[{\includegraphics[width=1in,height=1.294in]{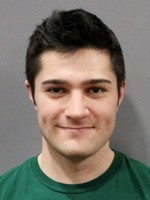}}]{Samuel C.~Chevalier} (S`13) received M.S. (2016) and B.S. (2015) degrees in Electrical Engineering from the University of Vermont, and he is currently pursuing the Ph.D. in Mechanical Engineering from the Massachusetts Institute of Technology (MIT). His research interests include power system stability and PMU applications.
\end{IEEEbiography}

\begin{IEEEbiography}[{\includegraphics[width=1in,height=1.294in]{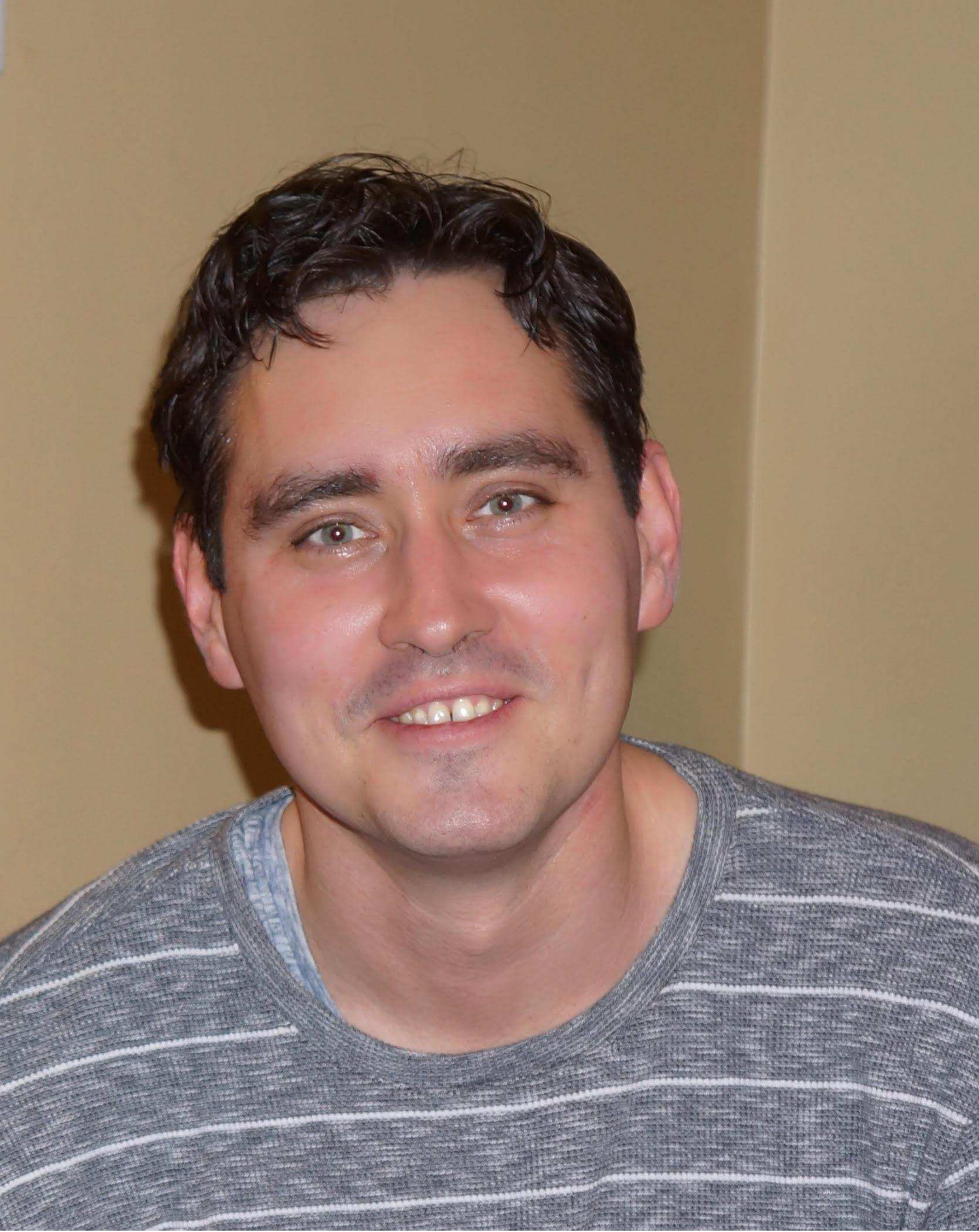}}] {Petr Vorobev}
(M`15) received his PhD degree in theoretical physics from Landau Institute for Theoretical Physics, Moscow, in 2010.  Currently, he is a Postdoctoral Associate at the Mechanical Engineering Department of Massachusetts Institute of Technology (MIT), Cambridge. His research interests include a broad range of topics related to power system dynamics and control. This covers low frequency oscillations in power systems, dynamics of power system components, multi-timescale approaches to power system modelling, development of plug-and-play control architectures for microgrids.
\end{IEEEbiography} 

\begin{IEEEbiography}[{\includegraphics[width=1in,height=1.294in]{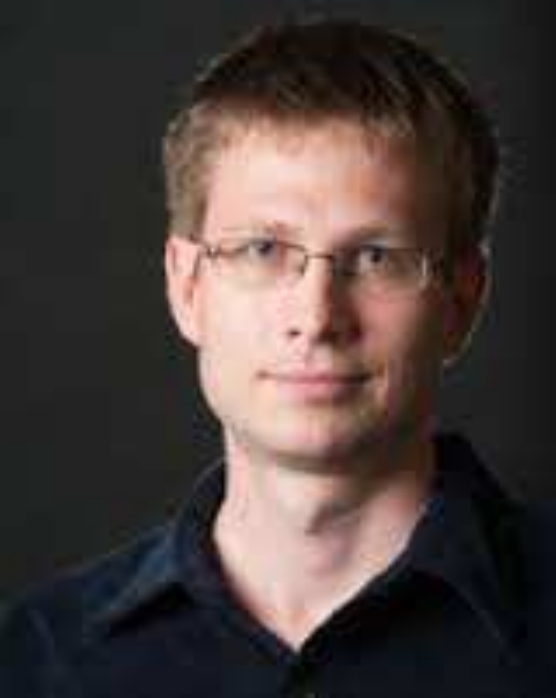}}]{Konstantin Turitsyn} (M`09) received the M.Sc. degree in physics from Moscow Institute of Physics and Technology and the Ph.D. degree in physics from Landau Institute for Theoretical Physics, Moscow, in 2007.  Currently, he is an Associate Professor at the Mechanical Engineering Department of Massachusetts Institute of Technology (MIT), Cambridge. Before joining MIT, he held the position of Oppenheimer fellow at Los Alamos National Laboratory, and Kadanoff-Rice Postdoctoral Scholar at University of Chicago. His research interests encompass a broad range of problems involving nonlinear and stochastic dynamics of complex systems. Specific interests in energy related fields include stability and security assessment, integration of distributed and renewable generation.
\end{IEEEbiography}
\end{document}